\documentclass[
aps, prd, reprint, notitlepage, a4paper,
floats, floatfix,
amsmath, amssymb, amsfonts, eqsecnum,
superscriptaddress,
showpacs, showkeys,
nofootinbib,
longbibliography,
]{revtex4-2}

\usepackage[left=20mm,right=20mm,top=25mm,bottom=25mm]{geometry}
\usepackage{graphicx} 
\usepackage[dvipsnames]{xcolor}
\usepackage{xspace} 
\usepackage{bm} 
\usepackage{amsmath,amsfonts,amssymb,amsthm}
\usepackage[utf8]{inputenc} 
\usepackage[normalem]{ulem}

\usepackage{overpic}
\usepackage{tikz}
\usepackage{float}

\xdefinecolor{customcolor}{rgb}{0.0, 0.53, 0.74}
\usepackage[
	bookmarksnumbered, bookmarksopen, bookmarksopenlevel=2,
	breaklinks=true,
	colorlinks=true, filecolor=customcolor, citecolor=customcolor,
	linkcolor=customcolor, urlcolor=customcolor, menucolor=customcolor,
]{hyperref}

\newcommand{\icasu}{\affiliation{The Grainger College of Engineering, Illinois Center for Advanced Studies of the Universe \& \\ Department of Physics, University of Illinois Urbana-Champaign, Urbana, IL 61801, USA}}
\newcommand{\UNLVphysics}{\affiliation{Department of Physics and Astronomy, University of Nevada, Las Vegas, 4505 South Maryland Parkway, Las Vegas, NV 89154, USA}}
\newcommand{\NCfA}{\affiliation{Nevada Center for Astrophysics, University of Nevada, Las Vegas, NV 89154, USA}}

\begin{document}

\title{Impact of numerical-relativity waveform calibration \\ on parametrized post-Einsteinian tests}

\hypersetup{pdftitle={Impact of numerical-relativity waveform calibration on parametrized post-Einsteinian tests}}

\author{Simone Mezzasoma}\email{simonem4@illinois.edu}\icasu
\author{Carl-Johan Haster}\email{carl.haster@unlv.edu}\UNLVphysics \NCfA
\author{Nicol\'as Yunes}\email{nyunes@illinois.edu}\icasu

\date{\today}

\begin{abstract}
Testing general relativity in the strong-field and highly dynamical regime is now possible through current gravitational-wave observations, where even a single high-quality detection can place competitive constraints on deviations from Einstein's theory. The parametrized post-Einsteinian framework provides a theory-agnostic approach to search for such deviations, but it typically assumes that systematic uncertainties in the base waveform model, particularly those arising from calibration to numerical relativity, are negligible. In this work, we investigate how calibration errors in the late-inspiral fitting coefficients of the IMRPhenomD waveform model can lead to spurious detections of departures from general relativity in parametrized tests. We use an uncertainty-aware version of IMRPhenomD, recalibrated to a set of numerical relativity surrogate waveforms and equipped with a probabilistic description of its fitting coefficients, to simulate general-relativity-consistent signals. We inject these signals into an O5 ground-based detector network and recover them with the original IMRPhenomD model augmented with a parametrized post-Einsteinian phase deformation. We find that false violations of general relativity using this model arise for network signal-to-noise ratios as low as 60. When the uncertainty-aware model is used instead, the inferred parametrized post-Einsteinian phase deformation remains consistent with zero even for signals with a signal-to-noise ratio up to 330. These results demonstrate the need to account for numerical relativity calibration uncertainty in order to perform reliable inspiral tests of general relativity. They also illustrate that explicitly incorporating numerical relativity calibration uncertainty into the waveform model preserves our ability to robustly test general relativity.
\end{abstract}

\maketitle

\section{Introduction}
\label{sec:introduction}

A decade of gravitational-wave (GW) detections since the first binary black hole (BBH) event, GW150914~\cite{LIGOScientific:2016aoc}, observed by the LIGO-Virgo-KAGRA (LVK) Collaboration~\cite{LIGOScientific:2014pky,VIRGO:2014yos,KAGRA:2020tym,LIGOScientific:2025slb-no-journal}, has enabled increasingly stringent tests of general relativity (GR) in the strong-field regime. 
While the Einstein field equations have been exceptionally successful at describing gravitational phenomena so far \cite{Will:2014kxa,Will:2018bme,LIGOScientific:2021sio}, they are also exceptionally difficult to solve because they are nonlinear and invariant under general coordinate transformations \cite{Arnowitt:1962hi}. Consequently, no general (i.e. without imposing any symmetry) analytic solution is known for the relativistic two-body problem \cite{Damour:2013hea-no-journal}.
To describe coalescing binaries across the inspiral, merger, and ringdown regimes, we instead rely on semianalytical waveform models that combine tractable, but perturbative, analytic approximations \cite{Blanchet:2013haa,Berti:2009kk} with full numerical relativity (NR) solutions obtained by discretizing the Einstein equations \cite{Gourgoulhon:2007ue-no-journal,Pretorius:2005gq,Campanelli:2005dd}.

Computationally efficient semianalytical waveform models are then central to detection, parameter estimation, and the ability to robustly test GR.
In constructing these models, whether phenomenological \cite{Ajith:2007kx,Khan:2015jqa,Hannam:2013oca,Khan:2018fmp,Pratten:2020ceb,Thompson:2023ase,Colleoni:2024knd,Hamilton:2025xru} or effective-one-body (EOB) \cite{Buonanno:1998gg,Buonanno:2009qa,Pan:2009wj,Bohe:2016gbl,Nagar:2018zoe,Pompili:2023tna,Ramos-Buades:2023ehm,Estelles:2025zah}, we face the theoretical challenge of ensuring that they faithfully represent the theory of gravity we set out to test.
The term \textit{waveform systematics} \cite{Cutler:2007mi,Vallisneri:2013rc,LIGOScientific:2016ebw} broadly describes the discrepancy between the operational GW model used in the analysis and the exact GW model that we would have obtained from the theory if we had unlimited computational power or more advanced theoretical tools.
The importance of waveform systematics on the robustness of tests of GR with GWs depends both on the source configuration and on how much of the signal falls within the sensitive band of our detectors.

The current ground-based detectors operate in observing runs, during which the instruments collect science data, interspersed with periods of construction and upgrades~\cite{prospectLVK_2020_version,Cahillane:2022pqm,Capote:2024rmo}. The LVK Collaboration has recently concluded the fourth observing run (O4)~\cite{Capote:2024rmo,LIGOScientific:2025snk-no-journal,LIGOScientific:2025slb-no-journal}, reporting the two loudest~\cite{LIGOScientific:2025cmm-no-journal,LIGOScientific:2025rid} and the most massive~\cite{LIGOScientific:2025rsn} BBH mergers to date. The detectors are now being enhanced in preparation for the fifth observing run (O5), which will be preceded by the six-month first intermediate run (IR1)~\cite{IGWN:ObservingPlans}.
In light of the O5-era upgrades and the future construction of LIGO-India \cite{Saleem:2021iwi,Shukla:2023kuj}, the Einstein Telescope~\cite{ET:2019dnz,Branchesi:2023mws,ET:2025xjr}, and Cosmic Explorer \cite{Reitze2019Cosmic,Evans:2021gyd-no-journal}, the need to control waveform systematics for comparable-mass compact binaries is increasingly pressing.
As network sensitivities improve, the signal-to-noise ratio (SNR) of future detections is bound to rise to the point where statistical uncertainties on inferred binary properties no longer dominate over waveform systematics, and the latter become consequential~\cite{Purrer:2019jcp}.
Aside from biasing the astrophysical source parameters, which can impact population inference \cite{Singh:2023aqh,Dhani:2024jja,Kapil:2024zdn}, inadequate waveform models can allow modeling errors within GR to be mistaken for genuine deviations from the theory \cite{Gupta:2024gun-non-inspire-journal}.

An arsenal of techniques have been developed to assess the consistency of detected GW signals with GR and to compare them against the predictions of competing alternative theories \cite{Yunes:2013dva,Carson:2020-review,Krishnendu:2021fga,Gupta:2025utd-no-journal}. Residual tests check whether the noisy detector data, after subtracting the best-fit signal, is statistically consistent with detector noise \cite{LIGOScientific:2016lio,LIGOScientific:2019fpa}. In the LVK Collaboration analyses, this is done by subtracting the inferred maximum-likelihood GR waveform, specifically modeled with \texttt{IMRPhenomXPHM}~\cite{Pratten:2020ceb} in the latest analysis of the O4 events~\cite{LIGOScientific:2026qni-no-journal}. Alternative implementations of residual tests employ NR waveforms \cite{Boyle:2019kee,Healy:2022wdn,Hamilton:2023qkv,Scheel:2025jct} when available, NR-based surrogate models \cite{Varma:2018mmi,Varma:2019csw} or model-agnostic wavelets \cite{Cornish:2014kda,Cornish:2020dwh}, though their performance hinges on the quality of the signal reconstruction \cite{Cornish:2011ys,Vallisneri:2012qq}. Spline-based tests \cite{Edelman:2020aqj,Read:2023hkv} allow for smoothly varying, but otherwise generic deviations from a baseline waveform in order to capture possible beyond-GR physics, including rapid phase changes like those predicted by spontaneous scalarization \cite{Sampson:2014qqa}. Inspiral-merger-ringdown (IMR) consistency tests \cite{Ghosh:2016qgn,LIGOScientific:2016lio,LIGOScientific:2017bnn,Ghosh:2017gfp} instead assess whether the parameters of the binary and its remnant, estimated from separate segments of the signal, are mutually consistent under GR. 
Non-GR effects affecting the propagation of GWs \cite{LIGOScientific:2025jau-no-journal, Johnston:2025qmo} have also been probed by testing the graviton dispersion relation \cite{Will:1997bb,Mirshekari:2011yq} and the dependence of the GW strain on the luminosity distance, which changes in theories with extra dimensions \cite{Deffayet:2007kf,Pardo:2018ipy}.
A more informative class of tests is represented by parametrized tests \cite{Yunes:2009ke,Cornish:2011ys,Arun:2006yw,Mishra:2010tp,Li:2011cg,Agathos:2013upa,Mehta:2022pcn}, which augment existing waveform models with additional parameters that quantify potential deviations from GR.

Among these tests, the parametrized post-Einsteinian (ppE) framework \cite{Yunes:2009ke,Chatziioannou:2012rf} applied to the inspiral regime introduces a dephasing into the frequency-domain waveform model to represent beyond-GR effects. This dephasing can, in the small-coupling limit and at leading post-Newtonian (PN) order, be mapped to specific alternative theories of gravity. Constraints on the ppE deformation parameter, obtained through Bayesian parameter estimation against detector data, can then be interpreted as a null test of GR or translated into bounds on the coupling constants of specific alternative theories \cite{Yunes:2016jcc}, provided the GR limit lies within the inferred credible region. 
Inspiral tests of this kind are routinely performed in GW analyses~\cite{LIGOScientific:2018dkp,LIGOScientific:2019fpa,LIGOScientific:2020tif,LIGOScientific:2021sio,LIGOScientific:2026qni-no-journal,LIGOScientific:2026fcf-no-journal,LIGOScientific:2026wpt-no-journal}, and current ground-based detectors are capable of meaningfully constraining inspiral phase deviations even with a single event \cite{LIGOScientific:2025wao}.

These constraints are only as reliable as the base waveform template used to recover them. For example, omitting spin precession in the base template when recovering highly precessing\footnote{Specifically, in the regime $\chi_p > 0.8$, where $\chi_p$ parametrizes the spin-precession effects sourced by the spin components orthogonal to the orbital angular momentum~\cite{Schmidt:2014iyl}.}, edge-on GR injections can bias ppE parameters at SNRs as low as $\sim30$ \cite{Chandramouli:2024vhw}. This bias is statistically significant as Bayesian model comparison tends to prefer the ppE model over the absence of ppE parameters, and this trend increases with SNR. 
Neglecting higher, and predominantly subdominant, harmonics also has an impact on ppE inference, though to a lesser degree than spin-induced orbital precession \cite{Chandramouli:2024vhw}, because higher harmonics primarily affect the waveform amplitude and are therefore weakly correlated to the ppE phase parameter.
A related modeling issue is the use of circularized recovery models on signals with residual eccentricity \cite{Saini:2022igm, Narayan:2023vhm}.
In particular, Ref.~\cite{Narayan:2023vhm} shows that LVK parametrized tests falsely indicate GR violations when NR injections with eccentricity $\sim0.05$ (at reference frequency of $17$ Hz) are analyzed with circular templates using the O4 detector network.

While many forthcoming GW observations will enable precision tests of GR, only a subset will be suitable to use given the current landscape of waveform systematics. In practice, a test of GR is applicable only for sources whose dynamics are faithfully captured by the recovery model. For instance, if the signature of a beyond-GR theory is directly coupled to the spins of the binary, then a waveform complete in the spin sector is a prerequisite for drawing scientific conclusions from highly-spinning binaries, if we want to be sure that the data correctly prefers the alternative theory over GR. These considerations have motivated the community to develop the latest generation of semianalytical waveform models that more faithfully capture BBH dynamics by incorporating higher-order modes, spin precession, and eccentricity.

A selection of the most recent models includes the circular-orbit precessing phenomenological waveforms \texttt{IMRPhenomXPHM}~\cite{Pratten:2020ceb} (used for parametrized tests in Ref.~\cite{LIGOScientific:2025wao}), and \texttt{PhenomXO4a} \cite{Thompson:2023ase}, together with the time-domain \texttt{IMRPhenomTPHM} \cite{Estelles:2021gvs}.
In the EOB family, we find \texttt{TEOBResumSP} \cite{Akcay:2020qrj,Gamba:2021ydi} and \texttt{SEOBNRv5PHM} \cite{Ramos-Buades:2023ehm} (with its spin-aligned variant \texttt{SEOBNRv5HM} \cite{Pompili:2023tna}, used for parametrized tests in Ref.~\cite{LIGOScientific:2025cmm-no-journal,LIGOScientific:2025wao}). For a comprehensive benchmark of their performance against NR, see Ref.~\cite{MacUilliam:2024oif}. Additionally, time-domain models including eccentricity, but restricted to spin-aligned configuration have been developed: \texttt{IMRPhenomTEHM} \cite{Planas:2025feq}, \texttt{TEOBResumS-Dal\'i} \cite{Nagar:2024dzj}, \texttt{SEOBNRv5EHM} \cite{Gamboa:2024hli}, and \texttt{SEOBNREHM} \cite{Liu:2021pkr}. 
Recent efforts to jointly model orbital eccentricity and spin precession \cite{Klein:2018ybm,Phukon:2019gfh,Ireland:2019tao,Klein:2021jtd-no-journal,Fumagalli:2023hde,Arredondo:2024nsl}, and the slowly growing number of NR simulations containing both effects~\cite{Lewis:2016lgx,Knapp:2024yww}, have yielded full IMR waveform descriptions, including \texttt{SEOBNREPHM} \cite{Liu:2023ldr} and the spin-precessing extension of \texttt{TEOBResumS-Dal\'i} \cite{Gamba:2024cvy,Albanesi:2025txj} (used for parametrized tests in Ref.~\cite{Chiaramello:2025bhi-no-journal}).

One feature shared by all these advanced models, regardless of the amount of orbital dynamics they promise to include, is that the analytical information of the adiabatic inspiral and of the post-merger ringdown ultimately needs to be complemented by NR simulations. 
In phenomenological frequency-domain models, the PN template valid at low frequencies is extended into the late inspiral by promoting it to a more flexible ansatz containing pseudo-PN coefficients. These coefficients are then fit to a set of training NR waveforms at astrophysical parameter values that the model is expected to interpolate. The choice of ansatz and the procedure used to fit it are guided by physical arguments (e.g., adopting an extended PN series, expected to be at least as accurate as forthcoming analytical higher-order PN results) and by the end goal of maximizing the faithfulness of the model to the NR training data.
Given that generating BBHs waveforms with NR currently takes days to months, $\sim 7-9$ orders of magnitude slower than millisecond semianalytical models, NR calibration for these models will continue to be indispensable for detection and parameter estimation. 

The step of calibrating to NR alone introduces several sources of uncertainty into a model: 
\begin{itemize}
    \item[(1)] the ansatz parametrization can be ill-suited to capture NR features at high frequencies. This means that the model can struggle to generalize NR data, regardless of the fitting strategy used to tune it. We can estimate this uncertainty by comparing the calibrated model with the NR validation set, though silent failures may persist in regions of parameter space poorly represented in that set.
    \item[(2)] the choice of fitting algorithm, often guided by the ansatz parametrization, dictates how well we can avoid underfitting and overfitting. Different algorithms have been used, including least-square minimization of the waveform phase difference \cite{Santamaria:2010yb}, hierarchical fit \cite{Jimenez-Forteza:2016oae} with minimization of the waveform residual using collocation points \cite{Husa:2015iqa,Khan:2015jqa,Pratten:2020fqn,Garcia-Quiros:2020qpx}, mismatch minimization \cite{Lam:2023oga}, and likelihood-based Bayesian fits \cite{Bohe:2016gbl,Pompili:2024yec,Bachhar:2024olc,Mezzasoma:2025moh}.
    \item[(3)] the NR training set, whose composition is often limited by the number of simulations made available by the NR community \cite{Healy:2022wdn,Ferguson:2023vta,Hamilton:2023qkv,Scheel:2025jct}, may be too sparse across parameter space (e.g. for mass ratios $\geq 8$) and typically only spans tens of inspiral cycles before merger \cite{LISAConsortiumWaveformWorkingGroup:2023arg}.
    \item[(4)] NR waveform uncertainty, the contribution to it at present dominated by truncation errors (approximating the fields and their gradients with a finite resolution) \cite{Bernuzzi:2011aq,Zlochower:2012fk,Jan:2023raq} and extrapolation errors (working with a finite integration domain) \cite{Reisswig:2009rx,Lousto:2010qx,Hinder:2013oqa}. The NR uncertainty can be estimated by generating multiple instances of the same waveform with varied numerical settings, and is practically quantified by the mismatch between the highest and second-highest resolution \cite{Blackman:2015pia,Mitman:2025tmj}. Although the accuracy of current NR waveforms is sufficient for the GW events detected to date~\cite{Ferguson:2020xnm}, the impact of truncation errors on parameter estimation at higher SNR is being investigated using pure-NR recovery waveforms \cite{Lange:2017wki,Jan:2023raq}. A prescription for NR resolution has also been established to avoid recovery bias \cite{Ferguson:2020xnm}, though the distinguishability assumption \cite{Lindblom:2010mh} behind it is considered to be overly conservative \cite{Chatziioannou:2017tdw,Purrer:2019jcp}.
\end{itemize}

Parameter estimation pipelines with semianalytical models have largely bypassed these uncertainties, partly because tuning these models is already challenging, and the inclusion of calibration errors adds another layer of complexity.
In the case of phenomenological models, this is reflected in the use of fixed point estimates for the fitting coefficients. 
Models that use point estimates for the fitting coefficients then yield a single realization of the ``true'' GR signal for a set of astrophysical binary parameters.

Following proposed methods for waveform uncertainty quantification and marginalization \cite{Moore:2014pda,Doctor:2017csx,Read:2023hkv}, phenomenological and EOB model construction recently have begun to propagate calibration uncertainties into downstream parameter estimation, using Bayesian inference \cite{Pompili:2024yec,Mezzasoma:2025moh} or Gaussian process regression \cite{Khan:2024whs,Bachhar:2024olc}. These enhanced models can generate multiple waveform realizations for a set of astrophysical binary parameters, all consistent with the assumption that GR, as captured by NR, is the correct theory of gravity. 

The impact of including NR calibration uncertainties on astrophysical parameter estimates has only recently begun to be explored~\cite{Owen:2023mid, Pompili:2024yec, Kumar:2025nwb}, and a systematic assessment of how they affect ppE tests is still largely missing in the literature. In this paper, we take a first step toward filling this gap by quantifying how modifying items (2) and (3) in the calibration of the late-inspiral fitting coefficients in the phenomenological \texttt{IMRPhenomD}~\cite{Husa:2015iqa,Khan:2015jqa} model can bias inspiral ppE tests. We also investigate whether promoting fixed fitting coefficients to additional model parameters that encode the NR calibration uncertainty in (4) preserves the ability to correctly identify GR-consistent waveforms in ppE tests.

To address the first question, we consider GR-consistent instances of a recently recalibrated \texttt{IMRPhenomD} model, dubbed “uncertainty-aware”, that also encodes NR calibration uncertainty. We inject these signals into a LIGO-Virgo three-detector network with projected O5 sensitivity curves, and recover them with the original \texttt{IMRPhenomD} model augmented with a ppE phase parameter. We scan the ppE parameter space by varying the PN order at which the ppE phase correction enters, covering PN orders from $-1.5$ to $3.5$. For each PN order, we determine the SNR threshold above which false-positive deviations from GR emerge in a single GW event, and estimate the corresponding Bayes factor. We perform this study for two binary configurations: a lighter system with total mass $20\, M_\odot$ and a heavier system with total mass $60\, M_\odot$. For a value of mass ratio $q = 2.3$ and moderately high antialigned spins $\chi_1 = \chi_2 = -0.6$, we find that the threshold SNRs can be as low as $\sim 60$ for the lighter system with lower PN-order corrections ($<0$ PN), and as low as $\sim 50$ for the heavier system with higher PN-order corrections ($>1.5$ PN). To address the second question, we employ the uncertainty-aware \texttt{IMRPhenomD} as the base model in the ppE test itself and find that the inferred ppE parameter remains consistent with zero even for signals at SNR $\sim 330$.

The remainder of the paper is organized as follows.
In Sec.~\ref{sec:methods}, we describe the waveform models used in this work, review the inspiral ppE deformation, and define the statistical tools to identify GR violations and to carry out model selection.
In Sec.~\ref{sec:analysis}, we present our injection-recovery study, report the threshold SNRs and Bayes factors associated with false-positive ppE inferences, and show that marginalizing over NR-calibration uncertainty restores consistency with GR.
We summarize our conclusions and outline future directions in Sec.~\ref{sec:conclusions}.
Throughout this work, we adopt geometric units with $G = c = 1$.

%%%%%%%%%%%%%%%%%%%%%%%%%%%%%%%%%%%%%%%%%%%%%%%%%%%%%%%%%%%%%%%%%%%%%%%%%%%
\section{Methods}
\label{sec:methods}

We begin by reviewing the \texttt{IMRPhenomD} model and its uncertainty-aware extension in Sec.~\ref{ssec:GR-waveforms}, which we adopt as our GR-consistent reference waveform model. In Sec.~\ref{ssec:beyond-GR-waveforms}, we summarize the ppE framework used to test GR, introduce the criterion for claiming a GR violation, and define an enhanced ppE model built on the uncertainty-aware \texttt{IMRPhenomD}.

\subsection{GR-consistent waveform}
\label{ssec:GR-waveforms}
Before assessing how calibration uncertainties in phenomenological waveform models affect inspiral parametrized tests of GR, we must define a representative GR waveform that serves as the baseline for our analysis. We select a circular, spin aligned, and near equal mass\footnote{Hence most of the signal power is carried by the dominant $\ell = 2 = |m|$ mode.} source configuration to represent the essential BBH dynamics in GR, while remaining simple enough so that we can easily interpret the effects of NR ``miscalibration'' in a realistic parameter estimation.

Hence, we adopt the \texttt{IMRPhenomD} model of~\cite{Husa:2015iqa,Khan:2015jqa} as our starting point for a baseline model consistent with GR. We could have instead used the more recent \texttt{IMRPhenomXAS} model~\cite{Pratten:2020fqn}\footnote{The \texttt{IMRPhenomXAS} model retains the key features of the \texttt{IMRPhenomD} model but provides a more effective parametrization of the spin sector, incorporates more recent PN results in the inspiral regime, and is trained on a larger set of NR simulations.} to model such sources, but we will instead employ the \texttt{IMRPhenomD} model with the understanding that the analysis described in Sec.~\ref{sec:analysis} can be extended to \texttt{IMRPhenomXAS} once a dedicated recalibration is performed. While the SNR thresholds at which false deviations from GR arise will differ from model to model, the main result in Sec.~\ref{sec:analysis} remains valid, i.e.~incorporating uncertainty in the fitting coefficients still allows robust tests of GR.
In the following, we summarize the components of the \texttt{IMRPhenomD} model that are relevant to this work, and refer the reader to Ref.~\cite{Husa:2015iqa} for a comprehensive description of the model.

The (frequency domain) phase of the \texttt{IMRPhenomD} model in the inspiral regime, defined for 
\begin{equation}
    Mf \in [0.0035, 0.018],
\end{equation} 
with $M = m_1 + m_2$ the total mass, can be expressed as a sum of two terms,
\begin{align}
\label{eq:GR-phase}
\phi_\mathrm{GR}(\boldsymbol{\theta},t_c,\phi_c, \boldsymbol{\lambda}; f) &=
\phi_{\mathrm{F2}}(\boldsymbol{\theta},t_c,\phi_c; f)
\\ \nonumber
&+ \phi_{\mathrm{cal}}(\boldsymbol{\theta},\boldsymbol{\lambda}; f) .
\end{align}
The phase evolution depends on the intrinsic parameters of the binary,
\begin{equation}
    \boldsymbol{\theta} = \lbrace m_1, m_2, \chi_1, \chi_2 \rbrace ,
    \label{eq:intrinsic-params}
\end{equation}
with $(m_1, m_2)$ ($m_1 \geq m_2$ by convention) the component masses and $(\chi_1,\chi_2)$ their respective dimensionless spins, assumed to be aligned with the orbital angular momentum. The model also depends on two arbitrary integration constants, $t_c$ and $\phi_c$, which are interpreted as time and phase at coalescence. 

The first term in Eq.~\eqref{eq:GR-phase}, $\phi_{\mathrm{F2}}$, combines known analytical PN results up to $3.5$ PN order. These are mapped from the time domain to the frequency domain under the stationary phase approximation~\cite{Thorne:1987,Cutler:1994ys,Poisson:1995ef,Droz:1999qx} (SPA) resulting in the \textit{canonical}\footnote{As opposed to \textit{extended}~\cite{Pratten:2020fqn,DuttaRoy:2024qbl}, which includes contributions up to $4.5$ PN order~\cite{Blanchet:2023sbv,Blanchet:2023bwj}.} TaylorF2 approximant~\cite{Buonanno:2009zt,Isoyama2020-review},
\begin{align}
\phi_{\mathrm{F2}}(\boldsymbol{\theta},t_c,\phi_c; f)
&= 2\pi f t_c - \phi_c - \frac{\pi}{4} \\ \nonumber
&+ \frac{3}{128} (\pi \mathcal{M} f)^{-5/3}
\sum_{i=0}^{7}
\varphi_i \, (\pi M f)^{i/3} ,
\end{align}
where the $i$th term in the sum is said to enter at  $i/2$ PN order.
The first four PN coefficients 
\begin{align}
\varphi_0 &= 1, \\
\varphi_1 &= 0, \\
\varphi_2 &= \frac{3715}{756} + \frac{55}{9}\eta , \\
\varphi_3 &= - 16\pi + \frac{113}{3} \chi_{\mathrm{PN}} ,
\end{align}
respectively introduce dependence on the chirp mass 
$\mathcal{M} = \eta^{3/5} M$ at $0$ PN order, 
the symmetric mass ratio $\eta = m_1 m_2 / M^2$ at $1$ PN order (or equivalently the mass ratio $q = m_1/m_2 \geq 1$, related to $\eta$ through $\eta = q/(1+q)^2$), 
and the effective spin parameter~\cite{Poisson:1995ef,Ajith:2011ec},
\begin{equation}
\chi_{\mathrm{PN}} = \frac{q \chi_1 + \chi_2}{1 + q} - \frac{38 \eta}{113} (\chi_1 + \chi_2),
\end{equation}
which first enters at $1.5$ PN order. The higher-order coefficients progressively reduce degeneracies among the intrinsic binary parameters and thus improve their measurability. The higher-PN order coefficients ($i \geq 4$) can be found in~\cite{Buonanno:2009zt,Isoyama2020-review}. 

The second term entering Eq.~\eqref{eq:GR-phase},
\begin{equation}
\label{eq:phi-cal}
\phi_{\mathrm{cal}}(\boldsymbol{\theta}, \boldsymbol{\lambda}; f) =
 \frac{3}{128} (\pi \mathcal{M} f)^{-5/3}
 \sum_{i=8}^{11} \kappa_i \, (\pi M f)^{i/3},
\end{equation}
is a collection of pseudo-PN terms that span from $4$ PN to $5.5$ PN order. These contain phenomenological coefficients
\begin{equation}
\label{eq:kappa}
\kappa_i = 
\frac{128}{\pi^{(i-5)/3}(i-5)} \, \sigma_{i-7},
\end{equation}
which depend on both the intrinsic parameters $\boldsymbol{\theta}$ and the NR-informed, late-inspiral fitting coefficients
\begin{equation}
    \boldsymbol{\lambda} = \lbrace \lambda^{j}_{00}, \lambda^{j}_{10}, \lambda^{j}_{01},\ldots, \lambda^j_{23}\rbrace\,,
\end{equation}
through the polynomial ansatz
\begin{align}
\label{eq:sigma}
\sigma_j &= \lambda_{00}^j + \lambda_{10}^j \eta \\ \nonumber
&\quad + (\chi_{\mathrm{PN}} - 1)
    \left( \lambda_{01}^j + \lambda_{11}^j \eta + \lambda_{21}^j \eta^2 \right) \\ \nonumber
&\quad + (\chi_{\mathrm{PN}} - 1)^2
    \left( \lambda_{02}^j + \lambda_{12}^j \eta + \lambda_{22}^j \eta^2 \right) \\ \nonumber
&\quad + (\chi_{\mathrm{PN}} - 1)^3
    \left( \lambda_{03}^j + \lambda_{13}^j \eta + \lambda_{23}^j \eta^2 \right).
\end{align}
The numerical prefactor in the definition of $\kappa_i$ in Eq.~\eqref{eq:kappa} is retained for continuity with the
original \texttt{IMRPhenomD} naming convention that used $\sigma_j$, but in practice it can be absorbed into the fitting coefficients $\boldsymbol{\lambda}$ upon rescaling them.

Several assumptions are made in choosing the ansatz in Eq.~\eqref{eq:phi-cal}. First, we only allow the pseudo-PN coefficients $\kappa_i$ to enter from $4$ PN order onwards, i.e.\ we set $\kappa_i = 0$ for $i < 8$. In practice, this amounts to treating the TaylorF2 approximant as PN-complete below $4$ PN order, even though this is known not to be true for spinning binaries. For example, at $3$ PN order the TaylorF2 phasing coefficient $\varphi_6$ contains only linear-in-spin corrections, and a non-zero $\kappa_6$ could in principle be used to model the missing quadratic and cubic spin contributions. We deliberately do not introduce such freedom and instead reserve the $\kappa_i$ for modeling truncation error at higher PN orders.
Second, we assume a purely power-series dependence on the orbital velocity $v = (\pi M f)^{1/3}$, even though the $4$ PN and $4.5$ PN phase coefficients are now known to contain non-analytic structures like $v^8 \ln v$,
$v^8 \ln^2v$, and $v^9 \ln v$, analogous to the $\ln v$ terms already present at $2.5$ PN and $3$ PN order in $\varphi_5$ and $\varphi_6$. The working assumption is that a PN-motivated power series with suitably chosen fitting coefficients at $4$ PN order and above can approximate these logarithmic contributions \textit{effectively}, thereby capturing the leading impact of PN truncation errors on the inspiral phase without explicitly resolving every higher-order term.

The calibration of the late-inspiral fitting coefficients (together with those
describing the intermediate and merger-ringdown segments) originally used 19 NR-EOB hybrid waveforms, concentrated along four mass ratio values
$q \in \{1,4,8,18\}$, with spin coverage
$\chi_{1,2} \in [-0.95,0.98]$ at $q=1$,
$[-0.75,0.75]$ at $q=4$, $[-0.85,0.85]$ at $q=8$, and $[-0.8,0.4]$ at $q=18$. Once optimized using a hierarchical fit, the coefficients were fixed to their best-fit values, here denoted as $\boldsymbol{\lambda}_\mathrm{PhenomD}$, which are implemented as explicit numerical constants in both \texttt{lalsuite}~\cite{lalsuite,swiglal} and \texttt{ripplegw}~\cite{Edwards:2023sak} libraries.

Fixing the fitting coefficients to numerical constants is standard practice when constructing semianalytical waveform models. This reflects the high level of care required when tuning them to NR data, as the maximally achievable faithfulness depends, not only on the way we perform the high-dimensional fit, but also on how we choose to represent the PN information. For example, adopting a different resummation scheme alone in EOB models can bring the EOB waveforms into significantly closer agreement with NR even without calibration~\cite{Damour:2007xr, Nagar:2016ayt, Messina:2018ghh}.
With new detector upgrades achieving detections with SNR on the order of $\sim 100$, GW data will place stringent demands on the accuracy of semianalytical models to the point where fixed fitting coefficients can become a limiting factor.

Work to embed NR calibration explicitly into waveform models has already begun for both EOB~\cite{Pompili:2024yec,Bachhar:2024olc} and phenomenological~\cite{Khan:2024whs,Read:2023hkv,Mezzasoma:2025moh} families, where the associated model uncertainties can be marginalized within a Bayesian framework. Complementary but related methods have also been developed to incorporate uncertainty due to inter-model discrepancy, directly in Bayesian inference~\cite{Hoy:2024vpc}. In the following, we use the uncertainty-aware extension of \texttt{IMRPhenomD}, introduced in previous work~\cite{Mezzasoma:2025moh}, as the waveform model that refines the hypothesis of GR being the correct theory by accounting for NR calibration uncertainty.

\begin{figure}
    \centering
    \includegraphics[width=\columnwidth]{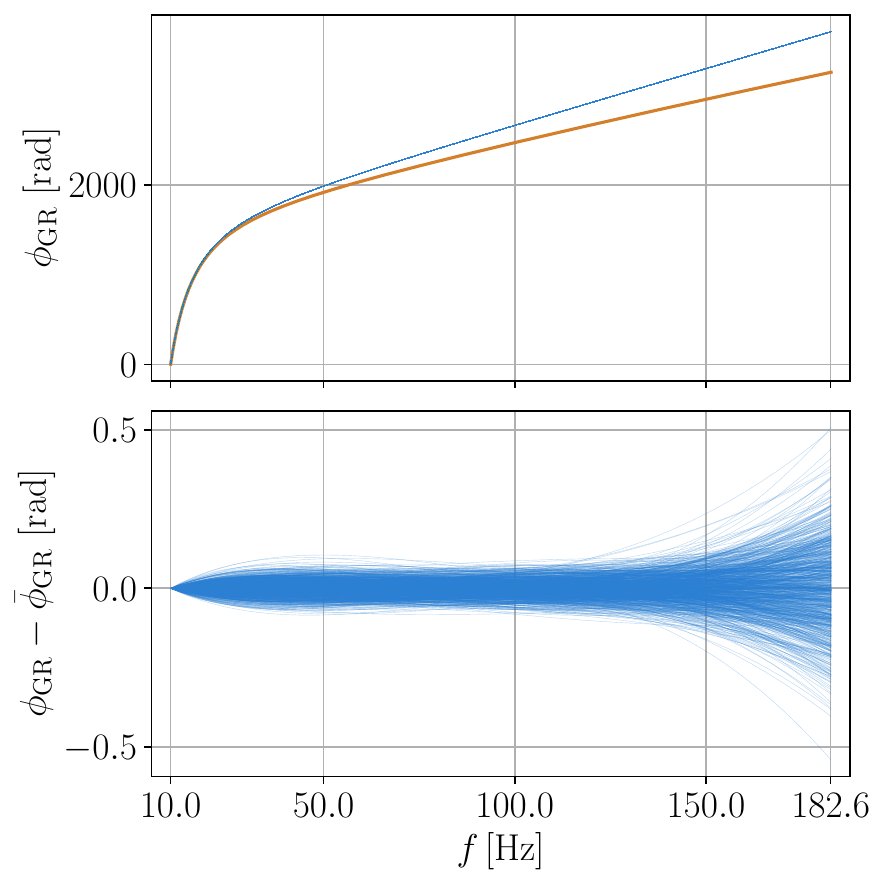}
    \caption{Phase evolution (top) and phase variability (bottom) for the lighter-mass binary system listed in
Table~\ref{tab:injection_params}, over the detector frequency range considered in this work. The top panel shows the GR phase
$\phi_{\mathrm{GR}}$, as
predicted by the vanilla \texttt{IMRPhenomD} model (orange) and its
uncertainty-aware recalibration (blue).
The bottom panel illustrates the variability of the phase residual $\phi_\mathrm{GR} - \bar{\phi}_{\mathrm{GR}}$,
obtained by drawing 1000 samples of fitting coefficients from
$\Pi(\boldsymbol{\lambda})$.}
    \label{fig:sampled_phase}
\end{figure}

In this model, the 33 fitting coefficients $\boldsymbol{\lambda}$ corresponding to $\sigma_2$, $\sigma_3$, and $\sigma_4$ in Eq.~\eqref{eq:sigma} (entering at $4.5$, $5$, and $5.5$ PN order, respectively), were jointly sampled in a Bayesian framework with a GW-analysis inspired likelihood, designed to match the model against a training set of 94 \texttt{NRHybSur3dq8}~\cite{Varma:2018mmi} waveforms. The coefficients entering $\sigma_1$ were instead kept fixed to their original values. 
The likelihood was evaluated over the inspiral range $0.0015 \leq Mf \leq 0.018$, and the training waveforms were distributed mostly uniformly in the parameter space $q\in [1,7.95]$, $\chi_{1,2} \in [-0.79, 0.79]$. The SNR scaling of each training waveform was chosen such that the mismatch (MM)~\cite{Cutler:1994ys,Owen:1995tm,Owen:1998dk}, computed over the posterior samples of the NR-calibration coefficients, would be centered around $10^{-4}$ (consistent with the typical minimum mismatch achievable by \texttt{IMRPhenomD}, as shown for example in Fig.~16 of Ref.~\cite{Pratten:2020fqn}) with a spread equal to $\sigma_\mathrm{MM} = 10^{-4}$. This choice of $\sigma_\mathrm{MM}$ effectively reproduces the typical uncertainty of \texttt{NRHybSur3dq8} (see e.g. Fig.~6 in Ref.\cite{Varma:2018mmi}), which is in turn used as a proxy for the underlying uncertainty of the NR simulations.
Observe that because the uncertainty-aware \texttt{IMRPhenomD} is trained on a different NR-based dataset and uses a different fitting procedure, it can differ significantly from the \texttt{lalsuite} version. The top panel of Fig.~\ref{fig:sampled_phase} shows that the phase difference reaches $\sim 400$ radians when we evaluate the two models on the lighter-mass binary parameters listed in Table~\ref{tab:injection_params}.

The posterior distribution of the fitting coefficients $\boldsymbol{\lambda}$ using the mismatch spread of $\sigma_\mathrm{MM} = 10^{-4}$ was found to be well approximated by a multivariate Gaussian (see insert in Fig. 2 of Ref.~\cite{Mezzasoma:2025moh}), which we denote here by $\Pi(\boldsymbol{\lambda})$. In what follows, we assume that the NR calibration uncertainty is fully described by the probability density $\Pi(\boldsymbol{\lambda})$, with $\bar{\boldsymbol{\lambda}}$ and $\boldsymbol{\Sigma}_\lambda$ denoting its mean vector and covariance matrix, respectively.
By modeling the fitting coefficients with a multivariate Gaussian distribution, the calibrated phase becomes a normally distributed quantity as well. We can see this by observing that the calibrated phase in Eq.~\eqref{eq:phi-cal} is linear in the fitting coefficients, and can therefore be written as
\begin{equation}
    \phi_{\mathrm{cal}}(\boldsymbol{\theta}, \boldsymbol{\lambda}; f) = \phi_{\mathrm{cal}}^{\sigma_1}(\boldsymbol{\theta}; f) + \sum_{k = 1}^{\dim (\boldsymbol{\lambda})} a_k (\boldsymbol{\theta}; f)\, \lambda_k
\end{equation}
where $\phi_{\mathrm{cal}}^{\sigma_1}(\boldsymbol{\theta}; f)$ denotes the contribution to the calibrated phase coming from $\sigma_1$, with its fitting coefficients fixed to their original \texttt{IMRPhenomD} values.
The coefficients $a_k(\boldsymbol{\theta}; f)$ can be obtained straightforwardly by taking the gradient with respect to the fitting coefficients,
\begin{equation}
a_k(\boldsymbol{\theta}; f) = \frac{\partial}{\partial \lambda_k} \phi_{\mathrm{cal}}(\boldsymbol{\theta}, \boldsymbol{\lambda}; f) .
\end{equation}
It then follows that, for fixed astrophysical parameters and GW frequency $f$, the phase $\phi_\mathrm{GR}(\boldsymbol{\theta},t_c,\phi_c, \boldsymbol{\lambda}; f)$ in Eq.~\eqref{eq:GR-phase} is a Gaussian random variable, induced by sampling the fitting coefficients $\boldsymbol{\lambda}$ from $\Pi(\boldsymbol{\lambda})$.
The mean and variance of $\phi_\mathrm{GR}(\boldsymbol{\theta},t_c,\phi_c, \boldsymbol{\lambda}; f)$ are then respectively given by
\begin{align}
\bar{\phi}_\mathrm{GR}(\boldsymbol{\theta},t_c,\phi_c; f) 
&= \phi_{\mathrm{GR}}(\boldsymbol{\theta},t_c,\phi_c, \bar{\boldsymbol{\lambda}}; f), \\
\sigma_{\phi_{\mathrm{GR}}}^2(\boldsymbol{\theta}; f) 
&= 
\boldsymbol{a}(\boldsymbol{\theta}; f)^\top
\boldsymbol{\Sigma}_\lambda \,\boldsymbol{a}(\boldsymbol{\theta}; f),
\end{align}
where the vector $\boldsymbol{a}(\boldsymbol{\theta}; f)$ has components $a_k(\boldsymbol{\theta}; f)$. 
The bottom panel of Fig.~\ref{fig:sampled_phase} shows a lightning plot of the phase variability, obtained by sampling the fitting coefficients from $\Pi(\boldsymbol{\lambda})$. For the lighter-mass binary parameters listed in Table~\ref{tab:injection_params}, the resulting phase spread of $\sim 0.25$ radians quantifies the NR calibration uncertainty that we marginalize over during parameter estimation in Sec.~\ref{sec:analysis}.

\subsection{Modeling NR-calibration uncertainty while testing GR}
\label{ssec:beyond-GR-waveforms}

In this work, we focus on inspiral-based tests of GR~\cite{Yunes:2013dva,Li:2011cg,Sampson:2013lpa,Agathos:2013upa,LIGOScientific:2019fpa,LIGOScientific:2021sio}. Focusing on the inspiral phase alone is well motivated even in the current detector era, as shown by the recently observed BBH event GW250114~\cite{Akyuz:2025seg-no-journal,LIGOScientific:2025rid} (SNR $\sim 65$ up to merger~\cite{LIGOScientific:2025wao}), which provides constraints on the inspiral comparable to, or more stringent than, those obtained by combining $\mathcal{O}(20)$ events from the fourth Gravitational-Wave Transient Catalog (GWTC-4.0)~\cite{LIGOScientific:2025slb-no-journal,LIGOScientific:2025wao}.

The ppE formalism \cite{Yunes:2009ke} provides a general framework for constructing inspiral waveform templates that deform a GR baseline model, such as \texttt{IMRPhenomD}, in a theory-agnostic way. These templates can then be used to test GR and quantify the evidence for alternative theories. 
Under the assumption that deviations from GR are controlled by a coupling constant that is much smaller than the length scale of the system we probe (a regime known as the small-coupling limit) \cite{Yunes:2009ke,Stein:2010pn,Yunes:2011we}, the leading beyond-GR contributions to the low-frequency GW signal can be computed perturbatively within the PN formalism. These can then be expressed in the frequency domain as corrections that scale as integer powers of the orbital velocity $v$.\footnote{Exceptions to this simple power-law behavior exist. These include non-adiabatic phenomena, like dynamical scalarization in scalar-tensor theories \cite{Barausse:2012da,Palenzuela:2013hsa,Shibata:2013pra,Taniguchi:2014fqa}, or non-analytical effects as found in GW emission sourced by a massive dark photon \cite{Alexander:2018qzg,Owen:2025odr}.}

Because GW detectors are more sensitive to variations in the phase than in the amplitude~\cite{Arun:2012hf,Tahura:2019dgr}, ppE modifications are most commonly applied to the phase alone. In their simplest form, these modifications introduce a single\footnote{For a study on multiparameter ppE tests, see Ref.~\cite{Perkins:2022fhr}.}, frequency-dependent dephasing term, so that the ppE phase becomes
\begin{align}
 \phi_{\mathrm{ppE}} (\boldsymbol{\theta},t_c,\phi_c, \boldsymbol{\lambda},\beta; f) &= \phi_{\mathrm{GR}}(\boldsymbol{\theta},t_c,\phi_c, \boldsymbol{\lambda}; f) \label{eq:ppE-phase} \\ \nonumber
 &+ \beta\, (\pi \mathcal{M} f)^{b/3} ,
\end{align}
with $\phi_{\mathrm{GR}}$ defined in Eq.~\eqref{eq:GR-phase}.
The deformation is characterized by (i) the exponent $b$, which is fixed \textit{a priori} in parameter estimation, and determines the PN order $(b+5)/2$ at which the non-GR correction enters the phase, and (ii) the amplitude parameter $\beta$, which is estimated from the data and quantifies the magnitude of the deviation from GR. This parameter serves as a phenomenological proxy for the coupling constants of the underlying theory, and reduces to the GR case in the limit $\beta = 0$.

To each value of the ppE exponent $b$ there corresponds one or more modified-gravity theories, in which distinct physical mechanisms produce phase corrections entering at that PN order \cite{Yunes:2016jcc}.
These mechanisms can act either in the generation of the signal, for example through additional radiating fields~\cite{Tahura:2018zuq}, or during its propagation, for example via a modified dispersion relation or Lorentz-violating effects that accumulate over cosmological distances \cite{Mirshekari:2011yq,Nishizawa:2017nef,LIGOScientific:2017bnn}.

A deviation with $b=-13$ ($-4$ PN order) can arise, for instance, in theories with a time-varying Newton's constant $G$~\cite{Yunes:2009bv,Tahura:2018zuq} or in phenomenological theories where gravity leaks into extra spatial dimensions \cite{Yagi:2011yu,Andriot:2017oaz}.
A deviation with $b = -7$ ($-1$ PN order) captures the effect of scalar dipole radiation. Such a term is predicted, for example, in Einstein-dilaton-Gauss-Bonnet \cite{Yagi:2011xp,Yagi:2012gp} and scalar-tensor~\cite{Horbatsch:2011ye,Jacobson:1999vr} theories.
A correction with $b = -5$ ($0$ PN order) is associated with the presence of a Lorentz-invariance-violating vector field, as in Einstein-Aether~\cite{Hansen:2014ewa} and khronometric theories~\cite{Hansen:2014ewa}. A deviation with $b = -3$ ($1$ PN order) corresponds to massive-graviton theories~\cite{Will:1997bb,Rubakov:2008nh,Hinterbichler:2011tt,deRham:2014zqa}. A deviation with $b = -1$ ($2$ PN order) is characteristic of dynamical Chern-Simons gravity \cite{Jackiw:2003pm,Alexander:2009tp,Yagi:2012vf}, a parity-violating extension of GR with a radiating pseudoscalar field. Finally, higher $b \geq 1$ ($\geq 3$ PN order) deviations can emerge in cubic and quartic gravity theories~\cite{Endlich:2017tqa,Cardoso:2018ptl,Liu:2024atc}, the gravitational Standard-Model Extension~\cite{Kostelecky:2016kfm}, or, in general, as next-to-leading PN order ppE corrections (as recently shown in purely metric quadratic gravity \cite{Alves:2025qcx}).

The ppE framework has been widely used to translate GW observations into constraints on the coupling constants of various modified-gravity theories since the first detection~\cite{Yunes:2016jcc,Nair:2019iur,Tan:2023fyl,Luo:2024vls}. In parallel, it has been extended to baseline waveform models that include additional polarizations~\cite{Chatziioannou:2012rf}, spin precession~\cite{Loutrel:2022xok}, and higher harmonics~\cite{Mezzasoma:2022pjb}. The robustness of ppE-inferred constraints when spin precession and higher harmonics are neglected in the recovery model has also been systematically investigated~\cite{Chandramouli:2024vhw}. More recently, an avatar of the ppE framework, realized through deep-learning neural networks~\cite{Xie:2024ubm,Xie:2025voe-no-journal}, has been developed to overcome the intrinsic limitation of fixing \textit{a priori} the exponent $b$, enabling a more general characterization of GR deviations.

To date, ppE constraints have relied on phenomenological waveform models whose calibration to NR simulations is treated as exact. In the specific case of the ppE model employing \texttt{IMRPhenomD}, the phase used for inference is
\begin{equation}
\phi_{\mathrm{ppE}}^{\mathrm{I}}(\boldsymbol{\theta},t_c,\phi_c,\beta; f)
= \phi_{\mathrm{ppE}}(\boldsymbol{\theta}, t_c,\phi_c,\boldsymbol{\lambda}_\mathrm{PhenomD},\beta; f) ,
\label{eq:model-I}
\end{equation}
with all phenomenological fitting coefficients fixed to their nominal calibrated value $\boldsymbol{\lambda}_\mathrm{PhenomD}$. We refer to this phase model as \emph{Model~I} henceforth.

When employed in null tests of GR, Model~I is exposed to any systematics inherited from the baseline model, among which those associated with its NR calibration. 
Before proceeding, we first specify our operational definition of a GR violation and then introduce the extended ppE waveform model.
Given a parameter-estimation run using Model~I, we assess a deviation from GR through the marginal posterior distribution $p(\beta)$ of the ppE amplitude parameter $\beta$. We define the $\gamma\%$ highest posterior density (HPD) region~\cite{Hyndman01051996,Gair-intro-to-stats-GW} as 
\begin{equation}
\label{eq:HPD}
    \mathcal{C}_{\gamma\%} = \left\{\, \beta : p(\beta) \geq c(\gamma) \,\right\},
\end{equation}
where $c(\gamma)$ is the largest constant such that  
\begin{equation}
\label{eq:HPD-90}
    \int_{\mathcal{C}_{\gamma\%}} p(\beta) \, d\beta = \gamma/100.
\end{equation}
We then declare a deviation from GR if the GR value $\beta = 0$ lies outside the $90\%$ HPD region,
\begin{equation}
    0 \notin \mathcal{C}_{90\%}.
\end{equation}
Using the HPD region provides a robust definition of GR violation even when the posterior for $\beta$ is multimodal, as can occur with \texttt{IMRPhenomD} for high-SNR signals from heavy binary systems. In the special case where the posterior for $\beta$ is symmetric and unimodal, the HPD region in Eq.~\eqref{eq:HPD} for $\gamma = 90$ coincides with the usual central credible interval bounded by the 5th and 95th percentiles of $\beta$.

We then introduce \emph{Model~II}, a generalization of Model~I that self-consistently incorporates NR calibration uncertainty through the uncertainty-aware version of \texttt{IMRPhenomD} as its baseline. In this case, the phenomenological fitting coefficients are no longer treated as fixed constants, but are instead sampled jointly with the astrophysical parameters.
The corresponding ppE phase model is
\begin{equation}
\phi_{\mathrm{ppE}}^{\mathrm{II}}(\boldsymbol{\theta},t_c,\phi_c,\boldsymbol{\lambda},\beta; f)
= \phi_{\mathrm{ppE}}(\boldsymbol{\theta}, t_c,\phi_c,\boldsymbol{\lambda},\beta; f) ,
\label{eq:model-II}
\end{equation}
where $\boldsymbol{\lambda}$ on the right-hand side is drawn from the multivariate Gaussian prior $\Pi(\boldsymbol{\lambda})$ defined in Sec.~\ref{ssec:GR-waveforms}. By comparing the performance of Models~I and II in recovering GR-consistent injections represented by an instance of the uncertainty-aware \texttt{IMRPhenomD} (with no ppE term), we can quantify in what circumstances a different calibration can mimic a beyond-GR effect, and assess whether the enhanced model is able to mitigate this bias.

In the following section, we motivate and define the specific GR-consistent signals used for the injection. 
We then describe the corresponding parameter estimation setup and results.

%%%%%%%%%%%%%%%%%%%%%%%%%%%%%%%%%%%%%%%%%%%%%%%%%%%%%%%%%%%%%%%%%%%%%%%%%%%
\section{Analysis and discussion}
\label{sec:analysis}

In this section we assess the impact of NR-calibration uncertainty on inspiral ppE tests by means of an injection-recovery study.
In Sec.~\ref{ssec:injection}, we describe our choice of BBH configuration and the NR-calibration fitting coefficients used for the injection.
In Sec.~\ref{ssec:parameter-estimation-pipeline}, we summarize the Bayesian parameter-estimation setup by defining likelihood, detector network, priors, and justifying the sampler settings. 
In Sec.~\ref{ssec:threshold-SNR-and-Bayes-factor}, we determine the threshold SNR beyond which we see false-positive GR violations using Model~I and compute the corresponding Bayes factor against pure GR.
In Sec.~\ref{ssec:recovery-at-high-SNR}, we recover the injections at SNR equal to $330$ to show that marginalizing over NR-calibration uncertainty restores consistency with $\beta=0$ in all PN orders we test.

%%%%%%%%%%%%%%%%%%%%%%%%%%%%%%
\subsection{Choice of source and NR-calibration fitting coefficients}
\label{ssec:injection}
In this work, we use the uncertainty-aware \texttt{IMRPhenomD} model reviewed at the end of Sec.~\ref{ssec:GR-waveforms} to generate signals that we treat as consistent with GR. Our aim is to identify plausible worst-case configurations in which NR-calibration systematics are most likely to bias the inferred ppE deformation parameter $\beta$. Quantifying the expected fraction of future detections impacted by such biases requires a dedicated population analysis and is left to future work.

To that end, we handpick two BBH configurations with moderately asymmetric masses and antialigned spins. We consider a \emph{lighter} system with total mass $M = 20\,M_{\odot}$ and a \emph{heavier} system with $M = 60\,M_{\odot}$. For both, we fix the mass ratio to $q = 2.303$ and set the dimensionless spins to $\chi_{1} = \chi_{2} = -0.6$. We focus on this region of intrinsic parameter space because that is where \texttt{IMRPhenomD} and the model used in our NR-calibration, \texttt{NRHybSur3dq8}, differ the most (see e.g. Fig.~3 of Ref.~\cite{Mezzasoma:2025moh}). We expect a ppE term to absorb these discrepancies and manifest as a bias in the inferred value of $\beta$ when the \texttt{IMRPhenomD} is ppE-enhanced.

As we broaden the interpretation of the GR hypothesis by considering any realization of the uncertainty-aware \texttt{IMRPhenomD}, we are free to select any set of fitting coefficients that is compatible with our NR calibration.
For our case study, we define a specific set of fitting coefficients, denoted by $\boldsymbol{\lambda}_\star$, picked from the 95\%-confidence ellipsoid of the distribution $\Pi(\boldsymbol{\lambda})$. 
Concretely, we define the 95\%-confidence ellipsoid 
\begin{equation}
   \mathcal{E}_{95\%} = \left\lbrace \boldsymbol{\lambda} : (\boldsymbol{\lambda}-\bar{\boldsymbol{\lambda}})^\top \boldsymbol{\Sigma}_\lambda^{-1}(\boldsymbol{\lambda}-\bar{\boldsymbol{\lambda}}) = F^{-1}_{\chi^2_{\nu}} (0.95) \right\rbrace ,
\end{equation}
where $F^{-1}_{\chi^2_{\nu}}$ is the quantile function of a chi-square distribution with 
$\nu = \dim(\boldsymbol{\lambda}) = 33$ degrees of freedom. 
The surface $\mathcal{E}_{95\%}$ represents the isoprobability surface enclosing 95\% of the samples of $\Pi(\boldsymbol{\lambda})$.

We choose $\boldsymbol{\lambda}_\star$ as the point on the surface $\mathcal{E}_{95\%}$ that maximizes the Euclidean distance from the nominal \texttt{IMRPhenomD} fitting coefficients $\boldsymbol{\lambda}_{\text{PhenomD}}$.
That is, we define 
\begin{equation}
\boldsymbol{\lambda}_\star 
= \underset{\boldsymbol{\lambda} \in \mathcal{E}_{95\%}}{\arg\max} 
\, \left\| \boldsymbol{\lambda} - \boldsymbol{\lambda}_{\text{PhenomD}} \right\|_2 ,
\end{equation}
where $\left\|\cdot  \right\|_2 $ denotes the Euclidean norm.

\begin{table*}
    \centering
    \begin{tabular}{lcc}
        \hline
        Source parameter & Lighter system & Heavier system \\
        \hline
        Total mass $M~[M_{\odot}]$ & 20.0 & 60.0 \\
        Chirp mass $\mathcal{M}~[M_{\odot}]$ & 7.87 & 23.60 \\
        Component masses $(m_1, m_2)~[M_{\odot}]$ & (13.95, 6.05) & (41.84, 18.16) \\
        Maximum frequency $f_{\mathrm{max}}~[\mathrm{Hz}]$ & 182.6 & 60.9 \\
        Minimum frequency $f_{\mathrm{min}}~[\mathrm{Hz}]$ & \multicolumn{2}{c}{10.0} \\
        Reference frequency $f_{\mathrm{ref}}~[\mathrm{Hz}]$ & \multicolumn{2}{c}{10.0} \\
        Mass ratio $q$ & \multicolumn{2}{c}{2.303} \\
        Spins $(\chi_1, \chi_2)$ & \multicolumn{2}{c}{(-0.6, -0.6)} \\
        Phase $\phi_{\mathrm{ref}}$ at $f_{\mathrm{ref}}$  $[\mathrm{rad}]$& \multicolumn{2}{c}{1.3} \\
        Polarization angle $\psi$ $[\mathrm{rad}]$ & \multicolumn{2}{c}{2.659} \\
        Geocentric time $t_{\mathrm{geo}}$ $[\mathrm{s}]$& \multicolumn{2}{c}{1126259642.413} \\
        Sky position $(\mathrm{ra}, \mathrm{dec})$ $[\mathrm{rad}]$ & \multicolumn{2}{c}{(1.375, -1.2108)} \\
        Inclination $\theta_{\mathrm{JN}}$ $[\mathrm{rad}]$& \multicolumn{2}{c}{0.4} \\
        \hline
    \end{tabular}
    \caption{Injection source parameters for the two BBH configurations employed in this study. Parameters that differ between the lighter and heavier systems are shown in separate columns, while parameters common to both configurations are listed below.}
    \label{tab:injection_params}
\end{table*}

%%%%%%%%%%%%%%%%%%%%%%%%%%%%%%
\subsection{Parameter estimation pipeline, model selection, and priors}
\label{ssec:parameter-estimation-pipeline}

Having defined a GW source to simulate and a waveform model to generate its signal, we now describe the Bayesian framework used to infer the properties of the binary~\cite{Cutler:1994ys,Rover:2006ni,LIGOScientific:2019hgc}, as well as any additional physics included in the recovery model, from the detector data $d(t)$ measured in response to the incident signal.

Under the hypothesis $\mathcal{H}$, which specifies a recovery model for the time-domain detector response function $h(\boldsymbol{\Xi},t)$, the signal is described by a set of parameters $\boldsymbol{\Xi}$. We infer these parameters by computing the posterior probability distribution $p(\boldsymbol{\Xi}| d,\mathcal{H})$ via Bayes' theorem~\cite{Gelman1995},
\begin{equation}
    p(\boldsymbol{\Xi} | d, \mathcal{H}) = \frac{\mathcal{L}(d | \boldsymbol{\Xi} , \mathcal{H}) \pi (\boldsymbol{\Xi} , \mathcal{H})}{Z(d | \mathcal{H})} ,
    \label{eq:bayes-theorem}
\end{equation}
where $\mathcal{L}(d | \boldsymbol{\Xi} , \mathcal{H})$ is the likelihood, namely the probability of observing the data $d(t)$ given the parameters $\boldsymbol{\Xi}$ and under  hypothesis $\mathcal{H}$, and $\pi (\boldsymbol{\Xi} , \mathcal{H})$ is the prior distribution encoding our assumptions about $\boldsymbol{\Xi}$ before analyzing the data. The evidence $Z(d | \mathcal{H})$, defined later in the text, serves as a normalization constant so the posterior integrates to unity, and it enables comparison between competing hypotheses.

For a single detector subject to stationary and Gaussian noise~\cite{Maggiore:2007ulw,creighton2011gravitational,Jaranowski:2005hz}, we adopt the Whittle \cite{whittle1951hypothesis,Cornish:2013nma,Thrane:2018qnx} likelihood,
\begin{equation}
    \mathcal{L}(d | \boldsymbol{\Xi} , \mathcal{H}) \propto e^{-\frac{1}{2} \langle d - h(\boldsymbol{\Xi}) | d - h(\boldsymbol{\Xi})  \rangle } ,
\end{equation}
where the noise-weighted inner product $\langle \cdot | \cdot \rangle$ between two time-domain signals $h_1(t)$ and $h_2(t)$ is defined by
\begin{equation}
    \langle h_1 | h_2 \rangle = 4\, \mathrm{Re} \int_{f_\text{min}}^{f_\text{max}} \frac{\tilde{h}_1(f) \tilde{h}^*_2(f)}{S_n(f)} df ,
    \label{eq:inner-product}
\end{equation}
with the one-sided noise power spectral density (PSD) $S_n(f)$. Here $\tilde{h}_1(f)$ is the Fourier transform of the time-domain signal $h_1(t)$, and the integration limits $[f_\mathrm{min} , f_\mathrm{max}]$ are defined by the detector bandwidth.

This motivates a choice of detector network. We select the detector configuration anticipated for the fifth observing (O5) run. This comprises the two Advanced LIGO detectors~\cite{LIGOScientific:2014pky}, Hanford (H1) and Livingston (L1), and the Advanced Virgo~\cite{VIRGO:2014yos} detector (V1), each performing at design sensitivity\footnote{The PSD of each detector can be found at \url{https://dcc.ligo.org/public/0165/T2000012/002/AplusDesign.txt} and at \url{https://dcc.ligo.org/public/0165/T2000012/002/avirgo_O5high_NEW.txt}.}~\cite{LIGO-T2000012}. Although KAGRA is also expected to join the O5 detector network, we do not include it here because its additional SNR contribution is expected to be modest relative to H1, L1, and V1, and its main benefits for parameter estimation are not central to the present study.
For a three-detector network generating multiple data streams $d = \lbrace d_D\rbrace_{D\in\{\mathrm{H1},\mathrm{L1},\mathrm{V1}\}}$, the overall likelihood factorizes as a product of the single-detector Whittle likelihoods,
\begin{equation}
\mathcal{L}( d | \boldsymbol{\Xi}, \mathcal{H} )
\propto 
\prod_{D\in 
\lbrace \mathrm{H1},\mathrm{L1},\mathrm{V1} \rbrace}
e^{
-\frac{1}{2}
\left\langle d_D - h_D(\boldsymbol{\Xi}) \,\middle|\, d_D - h_D(\boldsymbol{\Xi}) \right\rangle_D} ,
\end{equation}
where $\langle \cdot | \cdot \rangle_D$ is defined in Eq.~\eqref{eq:inner-product} evaluated using the PSD appropriate to detector $D$. Each detector contributes with an optimal, matched-filter SNR,
\begin{equation}
    \rho_D (\boldsymbol{\Xi}) = \sqrt{\langle h_D(\boldsymbol{\Xi}) | h_D(\boldsymbol{\Xi})\rangle_D} ,
\end{equation}
and these combine in quadrature to give the cumulative optimal network SNR, 
\begin{equation}
    \rho (\boldsymbol{\Xi}) = \sqrt{\sum_{D\in \lbrace \mathrm{H1},\mathrm{L1},\mathrm{V1}\rbrace} \rho_D^2 (\boldsymbol{\Xi}) } .
\end{equation}
For our O5-like network, we adopt a low-frequency cutoff of $f_{\min}=10\,\mathrm{Hz}$, consistent with the low-frequency sensitivity expected for O5. We set the high-frequency cutoff to $f_{\max}={0.018}/{M}$, where $M$ is the total mass of the injected signal, expressed in seconds. This choice restricts the likelihood to the inspiral portion of the signal, which is required because both the injected uncertainty-aware \texttt{IMRPhenomD} realization and the ppE recovery models used in this work are intended to be valid only in the inspiral regime; we therefore exclude merger and ringdown from the inference.
In all the injections we simulate, in each detector we pick a zero-noise realization for the noise PSD to characterize biases due to waveform mismodeling and distinguish them from fluctuations due to a specific noise realization.

Having specified the detector network, we now introduce the parameter set to be inferred.
With $\boldsymbol{\theta}$ the intrinsic parameters defined in Eq.~\eqref{eq:intrinsic-params}, we define the full set of 11 GR parameters as
\begin{equation}
    \boldsymbol{\Theta} = \boldsymbol{\theta} \cup \lbrace  t_\mathrm{geo} , \phi_\mathrm{ref}, \psi , \mathrm{ra}, \mathrm{dec}, \theta_\mathrm{JN}, D_L \rbrace ,
\end{equation}
which supplements the intrinsic parameters with the extrinsic parameters that determine the projection of the signal onto the detector network. Here $t_\mathrm{geo}$ is the arrival time at the geocenter (i.e. the coalescence time $t_c$ as measured from the Earth's center), and $\phi_\mathrm{ref}$ is a reparametrization of the orbital coalescence phase $\phi_c$, referenced to the frequency $f_\mathrm{ref}=10$ Hz. The polarization angle $\psi$ sets the orientation between the detector frame and the wave polarization plane, and $(\mathrm{ra},\mathrm{dec})$ specify the source sky location in right ascension and declination, which in turn define the detector pattern functions and the inter-detector time delays. The inclination $\theta_\mathrm{JN}$ is the angle between the total angular momentum and the line of sight, and it controls the relative contribution of the two GW polarizations. For more details on how these parameters are defined, see e.g. Table E1 of Ref.~\cite{Romero-Shaw:2020owr}. Finally, the luminosity distance $D_L$, inversely proportional to the amplitude $|h| \propto D_L^{-1}$, sets the scaling of the observed strain and thus is modified to inject the desired network SNR.

Note that the choice of $\phi_\mathrm{ref}, \psi, \mathrm{ra}, \mathrm{dec}$, and $\theta_\mathrm{JN}$ is not critical for our study, since our analysis employs a leading-order quadrupolar waveform model.
Because the NR fitting coefficients of the waveform amplitude are held fixed across the injection and all recovery models, the specific choice of these extrinsic parameters has no impact on the bias induced by our NR recalibration. The extrinsic parameters adopted for our injections (all except $D_L$, which we vary to span the network SNR range of interest), shown in Table~\ref{tab:injection_params}, are chosen so that the injected signal projects comparably onto H1, L1, and V1, and the individual detectors contribute roughly equally to the total network SNR, namely $\rho^2 /3 \approx \rho_\mathrm{H1}^2 \approx \rho_\mathrm{L1}^2 \approx \rho_\mathrm{V1}^2$.

In the following sections, we consider three recovery models, each associated with a working hypothesis. The baseline hypothesis $\mathcal{H}_0$ corresponds to the standard \texttt{IMRPhenomD} waveform, which is seen as a subset of the nested model $\mathcal{H}_{\mathrm{I}}$ when the ppE deformation is fixed to $\beta = 0$ (see the phase model in Eq.~\eqref{eq:model-I}). We then consider two beyond-GR models: $\mathcal{H}_{\mathrm{I}}$, which introduces the single ppE parameter $\beta$ and employs the phase in Eq.~\eqref{eq:model-I}, and $\mathcal{H}_{\mathrm{II}}$, which additionally includes the variable NR-calibration fitting coefficients $\boldsymbol{\lambda}$ as defined in Eq.~\eqref{eq:model-II}. For each hypothesis we will consider, the corresponding set of model parameters is:
\begin{align}
\mathcal{H}_0                &: \ \boldsymbol{\Xi} = \boldsymbol{\Theta},\\
\mathcal{H}_{\mathrm{I}}   &: \ \boldsymbol{\Xi} = \boldsymbol{\Theta} \cup \lbrace \beta \rbrace,\\
\mathcal{H}_{\mathrm{II}}  &: \ \boldsymbol{\Xi} = \boldsymbol{\Theta} \cup \boldsymbol{\lambda} \cup \lbrace \beta \rbrace.
\end{align}

A key feature of the Bayesian framework is that it allows one to marginalize over parameters that are not of direct interest, accounting for uncertainty in these parameters without fixing them to a particular value.
For example, in the case of $\mathcal{H}_\mathrm{I}$, we are interested in $\beta$ only. Its marginal posterior distribution is obtained by integrating over $\boldsymbol{\Theta}$,
\begin{equation}
    p(\beta) := p( \beta | d, \mathcal{H}_\mathrm{I}) = \int p(\boldsymbol{\Xi} | d, \mathcal{H}_\mathrm{I}) \, d \boldsymbol{\Theta} .
    \label{eq:marginal-beta}
\end{equation}
Similarly, in the case of $\mathcal{H}_\mathrm{II}$, we are interested in $\boldsymbol{\Theta}$ and $\beta$, while we regard the fitting coefficients $\boldsymbol{\lambda}$ as nuisance parameters. The marginal posterior for $\boldsymbol{\Theta} \cup \{\beta\}$ is obtained by integrating the joint posterior over $\boldsymbol{\lambda}$,
\begin{equation}
    p( \boldsymbol{\Theta}, \beta | d, \mathcal{H}_\mathrm{II}) = \int p(\boldsymbol{\Xi} | d, \mathcal{H}_\mathrm{II}) \, d \boldsymbol{\lambda} ,
\end{equation}
and this operation allows us to propagate the uncertainty encoded in the NR-calibration into the inferred astrophysical parameters and the beyond-GR coupling $\beta$.

The Bayesian evidence for a given hypothesis $\mathcal{H}$ in Eq.~\eqref{eq:bayes-theorem} is defined as the likelihood integrated over the full parameter space weighted by the prior,
\begin{equation}
    Z(d | \mathcal{H}) = \int \mathcal{L}(d | \boldsymbol{\Xi} , \mathcal{H}) \pi (\boldsymbol{\Xi} , \mathcal{H})\, d\boldsymbol{\Xi} .
    \label{eq:evidence}
\end{equation}
The evidence is primarily used for \textit{model selection}, where one compares two hypothesis and finds which one is statistically favored by the data \cite{Thrane:2018qnx}. 
As we are interested in quantifying the statistical significance of false-positive GR violations, we compare the two nested hypothesis $\mathcal{H}_\mathrm{I}$ (ppE) and $\mathcal{H}_0$ (pure GR). Model selection between the two is performed by computing the Bayes factor~\cite{Kass-Raftery}, defined as the ratio of the evidences,
\begin{equation}
\mathrm{BF}_{\mathrm{ppE},\mathrm{GR}} = \frac{Z(d | \mathcal{H}_{\mathrm{I}})}{Z(d | \mathcal{H}_0)} .
\label{eq:bayes-factor}
\end{equation}
In many cases, the Bayes factor can be smaller than the smallest floating-point number representable on a computer. Therefore, in practice to avoid underflow, it is common to work with the logarithm of the Bayes factor,
\begin{equation}
\log_{10}\mathrm{BF}_{\mathrm{ppE},\mathrm{GR}}
= \log_{10} Z(d | \mathcal{H}_{\mathrm{I}}) - \log_{10} Z(d | \mathcal{H}_0) .
\end{equation}
To interpret the value of  $\log_{10}\mathrm{BF}_{\mathrm{ppE},\mathrm{GR}}$, we follow Jeffreys' scale~\cite{Jeffreys:1939xee}, which we summarize in Table~\ref{tab:jeffrey-scale}. In general, negative values favor $\mathcal{H}_0$ over $\mathcal{H}_{\mathrm{I}}$, and the magnitude $\left|\log_{10}\mathrm{BF}_{\mathrm{ppE},\mathrm{GR}}\right|$ sets the strength of the evidence. A zero value means that the data is not able to prefer one model over the other. Conversely, a positive value favors $\mathcal{H}_{\mathrm{I}}$ over $\mathcal{H}_0$, with the same levels of evidence applying according to $\left|\log_{10}\mathrm{BF}_{\mathrm{ppE},\mathrm{GR}}\right|$. 
\begin{table}[!htbp]
\centering
\begin{tabular}{ll}
\hline
$\log_{10}\mathrm{BF}_{\mathrm{ppE},\mathrm{GR}}$ & Evidence level against $\mathcal{H}_{\mathrm{I}}$ \\
\hline
$< -2$ & Decisive \\
$[-2,\,-1.5)$ & Very strong \\
$[-1.5,\,-1)$ & Strong \\
$[-1,\,-0.5)$ & Substantial \\
$[-0.5,\,0)$ & Bare mention \\
\hline
\end{tabular}
\caption{Jeffreys' scale~\cite{Jeffreys:1939xee} used to interpret $\log_{10}\mathrm{BF}_{\mathrm{ppE},\mathrm{GR}}$ values, expressed as evidence against $\mathcal{H}_{\mathrm{I}}$ (ppE) in favor of $\mathcal{H}_0$ (GR).}
\label{tab:jeffrey-scale}
\end{table}

Parameter estimation is carried out with the \texttt{bilby} \cite{bilby_paper} library using its parallelized implementation, \texttt{parallel-bilby} \cite{pbilby_paper}. We use the \texttt{dynesty} \cite{2020MNRAS.493.3132S,sergey_koposov_2024_12537467} nested-sampling algorithm available in \texttt{bilby} to both sample the posterior probability in Eq.~\eqref{eq:bayes-theorem} and to compute the evidence in Eq.~\eqref{eq:evidence}, which is then used to obtain the Bayes factors in Eq.~\eqref{eq:bayes-factor}. Typical runs use \texttt{nlive} between $1500$ and $8000$ live points, \texttt{nact} between $2$ and $10$ autocorrelation lengths, and \texttt{walks} between $150$ and $600$ steps.
We adopt the higher end of these ranges to verify sampling convergence in special cases, where multiple neighboring modes are suspected near the maximum-likelihood region. The stopping criterion controlled by \texttt{dlogz} is set to either $0.1$ or $0.01$, the latter case when we are interested in computing the evidence with higher precision.

All waveforms are generated with the \texttt{ripplegw} \cite{ripple_paper} library, which we extend to include a JIT-compilable ppE phase modification as defined in Eq.~\eqref{eq:ppE-phase}. This infrastructure enables recovery with either the standard or the uncertainty-aware version of \texttt{IMRPhenomD}, by respectively using either a delta function $\delta^{(33)}(\boldsymbol{\lambda} -~\boldsymbol{\lambda}_\mathrm{PhenomD})$ or the multivariate Gaussian $\Pi (\boldsymbol{\lambda})$ as prior for $\boldsymbol{\lambda}$. The ppE phasing can formally be switched off by using a delta function $\delta(\beta)$, centered at $\beta = 0 $, as a prior for $\beta$.

We choose the duration $T$ of the injected data segment using the \texttt{lalsuite} routine \texttt{SimInspiralChirpTimeBound}, which provides an upper bound on the
inspiral time from the low-frequency cutoff to merger for a given binary configuration.
Denoting this estimate by $\tau$,
we choose a conservative duration that contains the full inspiral signal by rounding $\tau$ to the next power of two
\begin{equation}
T  = 2^{\left\lfloor \log_2 \tau \right\rfloor + 1} \,\mathrm{s},
\end{equation}
where $\left\lfloor \cdot \right\rfloor$ denotes the integer part.

Though \texttt{bilby} supports likelihood marginalization over several extrinsic parameters, we adopt different settings depending on the goal of the analysis. In particular, \texttt{bilby} can marginalize numerically over the luminosity distance $D_L$ using a precomputed lookup table~\cite{Singer:2015ema,Thrane:2018qnx}, marginalize numerically over the geocentric coalescence time $t_{\mathrm{geo}}$ using a method based on the Fast Fourier Transform (FFT)~\cite{Farr:2014FFT}, and marginalize analytically over the reference phase $\phi_{\mathrm{ref}}$~\cite{Veitchetal:2013phase}. Furthermore, \texttt{bilby} can reconstruct the posterior distributions of the marginalized parameters via dedicated algorithms post-sampling
\cite{Singer:2015ema,Singer:2016eax}.
While these options reduce the dimensionality of the parameter space, and thus, can aid convergence, they can also introduce resolution-dependent effects that are undesirable, if we require precise estimates of quantities, such as the threshold SNR separating GR-consistent and GR-inconsistent recoveries (as defined by the criterion in Sec.~\ref{ssec:beyond-GR-waveforms}).

In Sec.~\ref{ssec:threshold-SNR-and-Bayes-factor}, where we determine threshold SNR and compute Bayes factor at this SNR, we disable all marginalizations and instead sample explicitly over $D_L$, $\phi_{\mathrm{ref}}$, and $t_{\mathrm{geo}}$. We do not marginalize over $D_L$ to avoid loss of resolution due to the discretization of the distance lookup table. We also keep $\phi_{\mathrm{ref}}$ as an explicit sampling parameter, because we find that the phase-marginalized likelihood can produce slightly broader reconstructed posteriors, which could bias our estimate of the threshold SNR. Finally, we do not marginalize over $t_{\mathrm{geo}}$ because preliminary tests show that the FFT-based time marginalization introduces numerical artifacts, manifesting as a non-uniform posterior for the time jitter parameter. This sensitivity arises because the marginalization is performed on a discrete time grid with spacing $\Delta t = 1/f_s$, which can be too coarse for the FFT approximation to work. Although increasing $f_s$ would mitigate these artifacts, it would also cause the waveform-generation to outweigh the benefit of a faster convergence. We therefore find that directly sampling $t_{\mathrm{geo}}$ at $f_s=512$ Hz\footnote{Note that $f_s=512$ Hz implies a maximum frequency of $256$ Hz, which is sufficient for our analysis, since the inspiral of the lighter system ends at $182.6$ Hz.} is more efficient overall.

In Sec.~\ref{ssec:recovery-at-high-SNR}, our aim is instead to study the qualitative behavior of the uncertainty-aware recovery model $\mathcal{H}_{\mathrm{II}}$ at high SNR. In this regime, we do enable marginalization over $D_L$ and over $\phi_{\mathrm{ref}}$ to improve sampling efficiency, while continuing to sample $t_{\mathrm{geo}}$ explicitly for the same resolution-related reason described above.

As for priors of astrophysical parameters $\boldsymbol{\Theta}$, we choose those commonly employed in LVK analyses~\cite{LIGOScientific:2018mvr}. For sampling efficiency, we do not sample directly in the component masses but on the chirp mass $\mathcal{M}$ and the inverse mass ratio $1/q$ instead, the latter chosen instead of $q$ so the domain is bounded. We place a prior uniform in component masses on $\mathcal{M}$, centered at the injected value and, initially, $4 \sigma_\mathcal{M}$ wide, where $\sigma_\mathcal{M}$ is the projected standard deviation as estimated using the inverse of the Fisher information matrix~\cite{Finn:1992xs,Cutler:2007mi}. After a pilot parameter-estimation run, we adjust the prior width to ensure the posterior is fully contained within it and does not pile up against the boundaries of the prior. This adjustment is particularly necessary when the recovered chirp mass is biased, since the size of that bias is not known \textit{a priori}. When computing the Bayes factor in Sec.~\ref{ssec:threshold-SNR-and-Bayes-factor}, we use the same prior for chirp mass in the models we compare.

We also place a prior uniform in component masses on the inverse mass ratio, $1/q$, in the range $[1/20,1)$. The dimensionless spins $\chi_1$ and $\chi_2$ are taken to be uniform in the range $[-0.99,\,0.99]$. 
We use a uniform prior on the geocentric time $t_{\mathrm{geo}}$, centered on the injected value with a total width of $0.2$ s.
We implement an isotropic sky-location prior in equatorial coordinates: ra and $\cos(\mathrm{dec})$ are taken to be uniform with ra in $[0,2\pi)$ and dec in $[-\pi/2 , \pi/ 2]$. This assumes that the signal could be coming from any direction with uniform probability over the celestial sphere. We also employ a uniform prior for  $\sin\theta_{\mathrm{JN}}$ with $\theta_{\mathrm{JN}}$ in $[0,\pi ]$. We take a uniform prior for $\psi$ in $[0,\pi)$, and a uniform prior for $\phi_\mathrm{ref}$ in $[0,2\pi)$, both with periodic boundary condition.
Finally, the prior on the luminosity distance $D_L$ corresponds to a redshift distribution that is uniform in comoving volume and in source-frame time~\cite{Romero-Shaw:2020owr}. This reflects the assumption that merger events occur uniformly throughout the observable Universe.

We take the ppE parameter $\beta$ to have a uniform prior in the interval $[\beta_\mathrm{min},\beta_\mathrm{max}]$, where the bounds depend on the PN order (i.e.~$b$) and are chosen to satisfy two requirements.
First, the prior should maintain the physical validity of the ppE framework. This means it should ensure that the ppE phasing remains subdominant to the GR phasing over the detector band~\cite{Perkins:2022fhr}\footnote{An alternative is to choose a prior range informed by the tightest existing observational constraint on the coupling constants of modified-gravity theories that map to the chosen ppE exponent $b$~\cite{Xie:2024ubm}. We do not adopt this strategy so that our conclusion does not depend on constraints derived from previous events.}.
Second, the bounds should be tight enough to preserve a good resolution of the posterior, given a finite number of $\mathcal{O}(10^4)$ posterior samples. This is because an overly broad prior can spread the accepted samples thinly in the region supported by the likelihood, and this both increases the run time and can degrade the estimate of the evidence.

In practice, we initialize $[\beta_{\min},\beta_{\max}]$ using a Fisher-information-matrix estimate, then perform a pilot parameter-estimation run and iteratively expand the interval only as needed until it fully contains the tails of the posterior. At the SNRs we probe, these bounds remain within the perturbative regime, where the ppE framework is expected to be valid.
Since the Bayes factor depends on the choice of priors, we list the prior bounds for $\beta$ explicitly in Table~\ref{tab:beta-prior}, obtained with the procedure described above.

These bounds maximize the Bayes factor with respect to the choice of uniform prior that keeps the posterior unchanged. To see this, observe that by construction the integration domain of Eq.~\eqref{eq:evidence} along $\beta$, where the integrand is non-zero, is dictated by the likelihood because we work in the likelihood-dominated regime. Then, the Bayes factor computed using an interval $[\beta_\mathrm{min}, \beta_\mathrm{max}]$, of length $\Delta \beta = \beta_\mathrm{max} - \beta_\mathrm{min}$, is greater than the Bayes factor computed using a wider interval $[\beta_\mathrm{min}', \beta_\mathrm{max}'] \supset [\beta_\mathrm{min}, \beta_\mathrm{max}]$, of length $\Delta \beta'= \beta_\mathrm{max}' - \beta_\mathrm{min}'$, because
\begin{equation}
\log_{10}\mathrm{BF}'_{\mathrm{ppE},\mathrm{GR}}
=
\log_{10}\mathrm{BF}_{\mathrm{ppE},\mathrm{GR}}
+
\log_{10}\!\left(\frac{\Delta\beta}{\Delta\beta'}\right).
\end{equation}
This relation also allows us to straightforwardly rescale our reported Bayes factors to any alternative choice of uniform prior bounds.

\begin{table}[!htbp]
\centering
\begin{tabular}{c c c}
\hline
Order & Lighter system & Heavier system \\
\hline
$-1.5$ PN & $[-3.0,4.67]\!\times\! 10^{-7}$ & $[-4.54,9.46]\!\times \!10^{-6}$ \\
$-1$ PN& $[-3.41, 7.07]\!\times \!10^{-6}$ & $[-5.13,13.24]\!\times\! 10^{-5}$ \\
$-0.5$ PN & $[-5.69, 17.39]\!\times\! 10^{-5}$ & $[-6.38,10.78]\!\times\! 10^{-4}$ \\
$0$ PN  & $[-3.27, 4.55]\!\times\! 10^{-3}$   & $[-6.01,5.70]\!\times\! 10^{-3}$ \\
$0.5$ PN  & $[-1.83, 1.17]\!\times\! 10^{-2}$   & $[-3.82,2.35]\!\times\! 10^{-2}$ \\
$1$ PN  & $[-13.5,7.01]\!\times\! 10^{-2}$   & $[-2.04,1.08]\!\times\! 10^{-1}$ \\
$1.5$ PN  & $[-18.9,6.20]\!\times\!10^{-1}$   & $[-18.7,6.43] \!\times\! 10^{-1}$ \\
$2$ PN  & $[-11.40,7.13]\!\times\!10^{0}$   & $[-2.93,1.14] \!\times\! 10^1$ \\
$3$ PN  & $[-8.46,12.85]\!\times\! 10^1$   & $[-3.66,4.36] \!\times\! 10^{2}$ \\
$3.5$ PN  & $[-3.96,5.33]\!\times\! 10^{2}$   & $[-7.39, 7.93]\!\times\! 10^{2}$ \\
\hline
\end{tabular}
\caption{Prior interval $[\beta_{\min},\,\beta_{\max}]$ on the ppE parameter $\beta$ for each PN order explored in Sec.~\ref{ssec:threshold-SNR-and-Bayes-factor}, shown separately for the lighter (central column) and heavier (rightmost column) BBH systems.}
\label{tab:beta-prior}
\end{table}

%%%%%%%%%%%%%%%%%%%%%%%%%%%%%%
\subsection{Results for threshold SNR and Bayes factor}
\label{ssec:threshold-SNR-and-Bayes-factor}

In this subsection we quantify the network SNR at which Model~I ($\mathcal{H}_{\mathrm{I}}$) begins to yield a false-positive GR violation when recovering the injection described in Sec.~\ref{ssec:injection}. For different values of $b$ corresponding to the PN orders $\lbrace -1.5, 1, -0.5, \ldots , 3.5\rbrace$, we vary the luminosity distance to sweep the injected network SNR $\rho$ and perform parameter estimation with \(\mathcal{H}_{\mathrm{I}}\). We here define the \emph{threshold SNR}, $\rho_\mathrm{th}$, as the smallest $\rho$ for which the GR value $\beta=0$ falls outside the $90\%$ HPD region (defined in Sec.~\ref{ssec:beyond-GR-waveforms}) of the marginal posterior $p(\beta)$ in Eq.~\eqref{eq:marginal-beta}. The details on the procedure used to estimate $\rho_\mathrm{th}$ can be found in Appendix~\ref{sec:appendix-bracketing}.

\begin{figure}
    \includegraphics[width=\columnwidth]{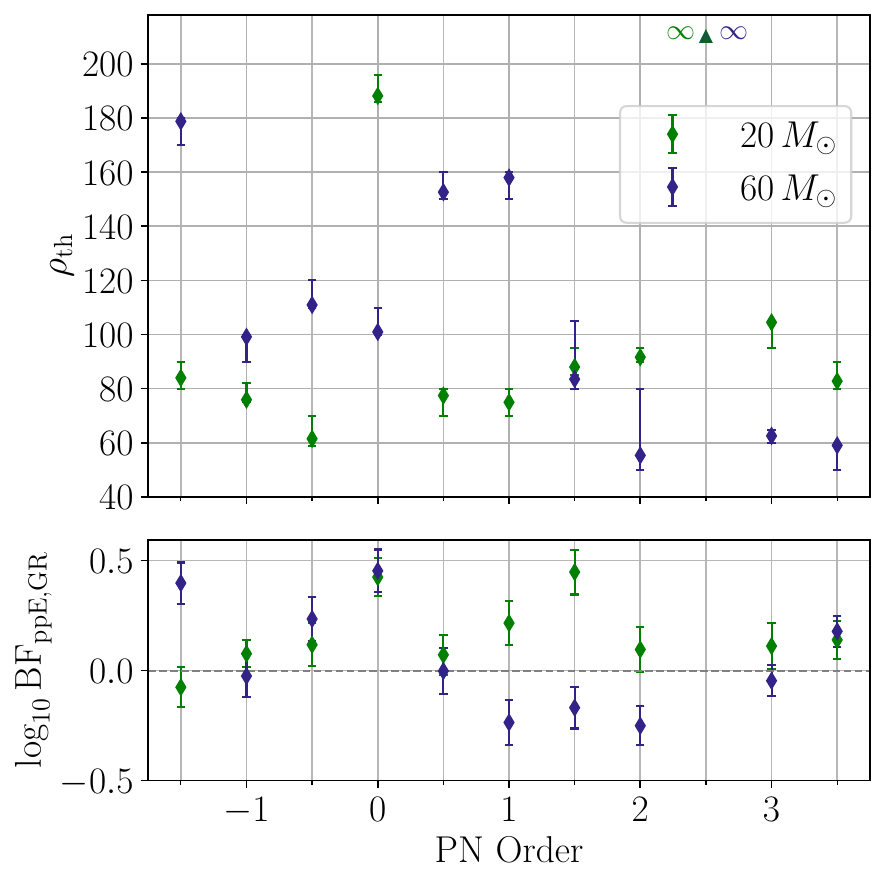}
    \caption{Upper panel: threshold network SNR, $\rho_{\mathrm{th}}$, as a function of the PN order $(b+5)/2$ at which the ppE dephasing enters, shown for the lighter ($20\,M_\odot$, green) and heavier ($60\,M_\odot$, blue) BBH systems. Diamond markers indicate the best estimate of $\rho_{\mathrm{th}}$, and the bracketing runs at slightly higher and lower SNR provide an uncertainty interval. At all PN orders we find $\rho_{\mathrm{th}}\gtrsim 60$. Lower panel: $\log_{10}\mathrm{BF}_{\mathrm{ppE},\mathrm{GR}}$ evaluated at the corresponding $\rho_{\mathrm{th}}$, with the associated uncertainty. The Bayes factors remain small in magnitude, $|\log_{10}\mathrm{BF}_{\mathrm{ppE},\mathrm{GR}}|\lesssim 0.5$, so while model selection at $\rho=\rho_{\mathrm{th}}$ is inconclusive, it also does not rule out a beyond-GR signature.}
    \label{fig:threshold_snr}
\end{figure}

The upper panel of Fig.~\ref{fig:threshold_snr} shows that the threshold SNR depends strongly on the PN order and on the total mass. This means that the phase residual induced by the NR-calibration projects differently onto the ppE basis as the number of cycles in band changes. In particular, false GR violations appear at network SNRs as low as $\rho \sim 60$ for the lighter system ($20\,M_\odot$) at $-0.5$ PN order, and as low as $\rho \sim 50$ for the heavier system ($60\,M_\odot$) at 2 PN order.

We find that formally $\rho_{\mathrm{th}}\sim \infty$ when the ppE correction enters at $2.5$ PN order. At this order, the ppE dephasing becomes completely degenerate with the phase at coalescence, which is equipped with a uniform prior with periodic boundary condition. As the quadrupole-only waveform model we use is unable to measure the phase at coalescence, this degeneracy prevents the data from placing a meaningful constraint on $\beta$.

We keep the uncertainty on the threshold estimates at SNRs within $\sim 10$ by bracketing the threshold from above and below with multiple runs. The only exceptions, with uncertainties of $\sim 30$, occur for the heavier system at $1.5$ and $2$ PN order. For these, as the SNR increases, the marginal posterior tightens predominantly on one side and this limits the precision with which the crossing of $\beta = 0$ can be identified.
This behavior arises because $\beta$ can slide along the correlations with the astrophysical parameters whose marginal posteriors are still noticeably non-Gaussian. In fact, although the marginal posterior distributions of each parameter are expected to become Gaussian in the $\rho\to\infty$ limit \cite{Flanagan:1997kp,Cutler:2007mi}\footnote{Assuming there is only one maximum-likelihood point, this remains true even for non-uniform priors: when $\rho$ is large, the likelihood concentrates into a very small region in which the prior can be considered constant.}, they do so at different rates which causes the asymmetric contraction of $p(\beta)$.

For the lighter system, $\rho_{\mathrm{th}}$ exhibits a pronounced spike at $0$ PN order relative to the neighboring PN orders, consistent with the strong degeneracy between the ppE parameter $\beta$ and the chirp mass $\mathcal{M}$. At this order, both $\beta$ and $\mathcal{M}$ control the leading-order phase evolution, and the degeneracy causes their marginal posterior to broaden significantly. The weaker effective constraint on $\beta$ requires a larger SNR before the $90\%$ HPD region excludes $\beta=0$.

We also observe a comparable increase in $\rho_{\mathrm{th}}$ at $0.5$ and $1$ PN order for the heavier system, compared to the neighboring PN orders. This is likely caused by (1) the marginal posterior for $\beta$ being bimodal at these orders, consistent with \texttt{IMRPhenomD} admitting multiple maximum-likelihood combinations of $\beta$ and $\mathbf{\Theta}$, and (2) this particular binary configuration allowing the phase residual to be more easily absorbed by shifts in the astrophysical parameters rather than a shift in $\beta$ at $0.5$ and $1$ PN order.

The qualitative trends of $\rho_\mathrm{th}$ in the upper panel of Fig.~\ref{fig:threshold_snr} can be understood heuristically by recalling how the ppE dephasing accumulates across the detector band. For the lighter binary, a low-PN-order correction builds up primarily at low frequency, where the signal remains in band for many cycles and contributes to a large fraction of the accumulated SNR. These condition allows to constrain the ppE term well, so the posterior $p(\beta)$ becomes sufficiently narrow that a false GR violation is reached at comparatively low $\rho_{\mathrm{th}}$. By contrast, a high-PN-order correction activates mostly at high frequency. For the lighter system, the inspiral carries a smaller fraction of the total SNR at high frequency, so a much louder injection is required to tighten $p(\beta)$ enough to exclude $\beta = 0$. The heavier system exhibits the complementary behavior, because a larger fraction of its SNR is concentrated in the high-frequency portion of the inspiral.

To assess whether these false-positive violations are statistically significant, we perform a second parameter estimation to compute the Bayes factor $\mathrm{BF}_{\mathrm{ppE},\mathrm{GR}}$ between $\mathcal{H}_{\mathrm{I}}$ and $\mathcal{H}_0$ at the corresponding $\rho_\mathrm{th}$, shown in the lower panel of Fig.~\ref{fig:threshold_snr}. The Bayes factor reveals which model is more likely to have produced the observed data and combines the goodness of fit provided by the likelihood with the Occam penalty associated with the prior volume on $\beta$~\cite{Trotta:2008qt}. In our case, it helps us answer the question: how convincingly can NR ``miscalibration'' mimic a GR violation? 

For the lighter system, the $\log_{10}\mathrm{BF}_{\mathrm{ppE},\mathrm{GR}}$ values are systematically positive, indicating a mild preference for the ppE hypothesis, with the exception of the $-1.5$-PN-order case. The lower value at $-1.5$ PN order can be explained by the frequency dependence of the phase residual between injection and recovery, caused by different values in the NR-calibration fitting coefficients. As the residual dephasing is primarily accumulated at higher frequencies, it more closely resembles a high-PN-order correction and a negative PN order ppE deformation provides little improvement in the fit\footnote{The Bayes factor is then dominated by the Occam penalty associated with the additional prior volume of $\beta$.}. In all cases, however, the model selection remains inconclusive, since the magnitude of $\log_{10}\mathrm{BF}_{\mathrm{ppE},\mathrm{GR}}$ remains within $\lesssim 0.5$ (see Table~\ref{tab:jeffrey-scale} for reference).

\begin{figure*}[htb]
    \includegraphics[width=\columnwidth]{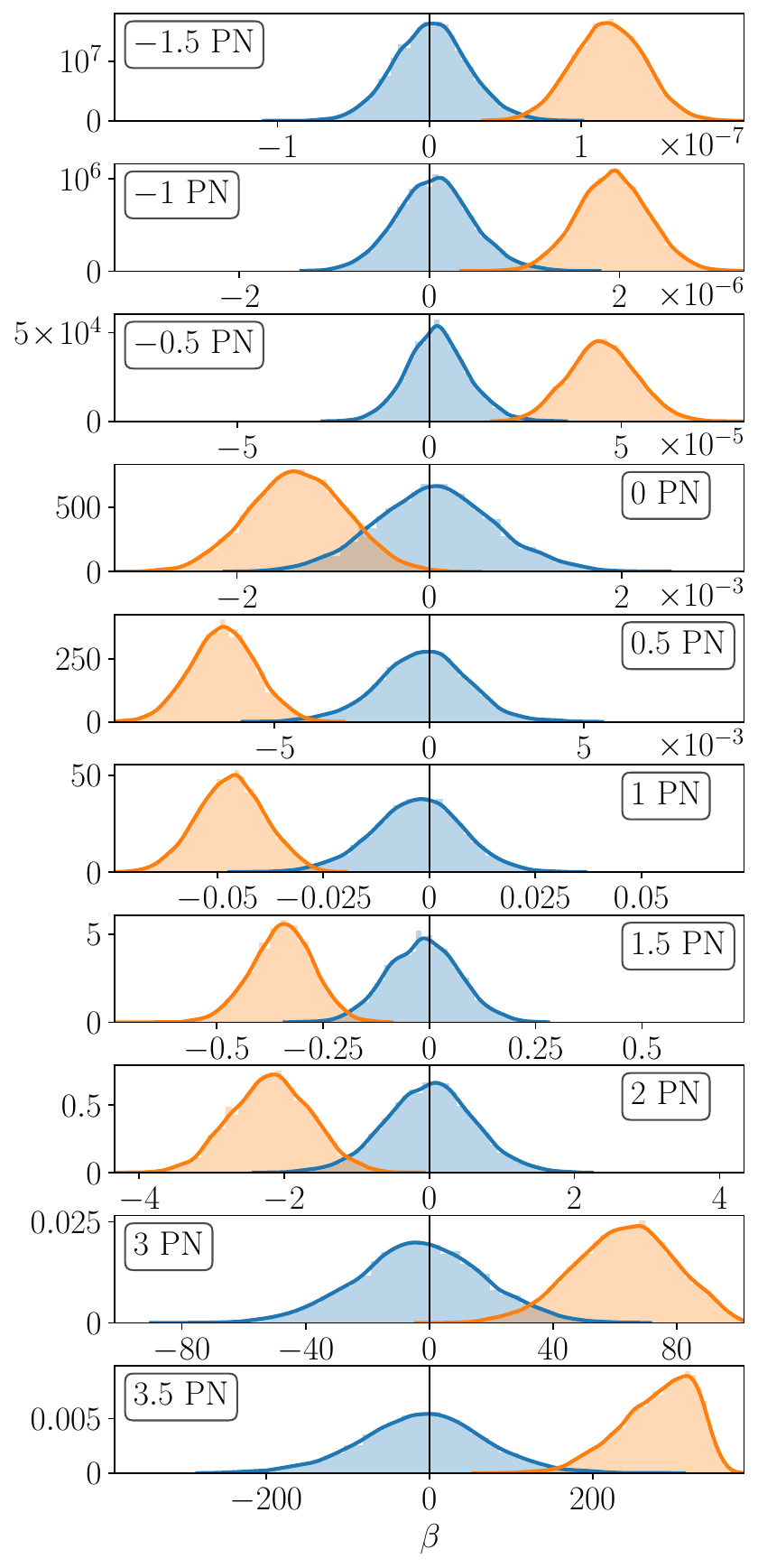}
    \includegraphics[width=\columnwidth]{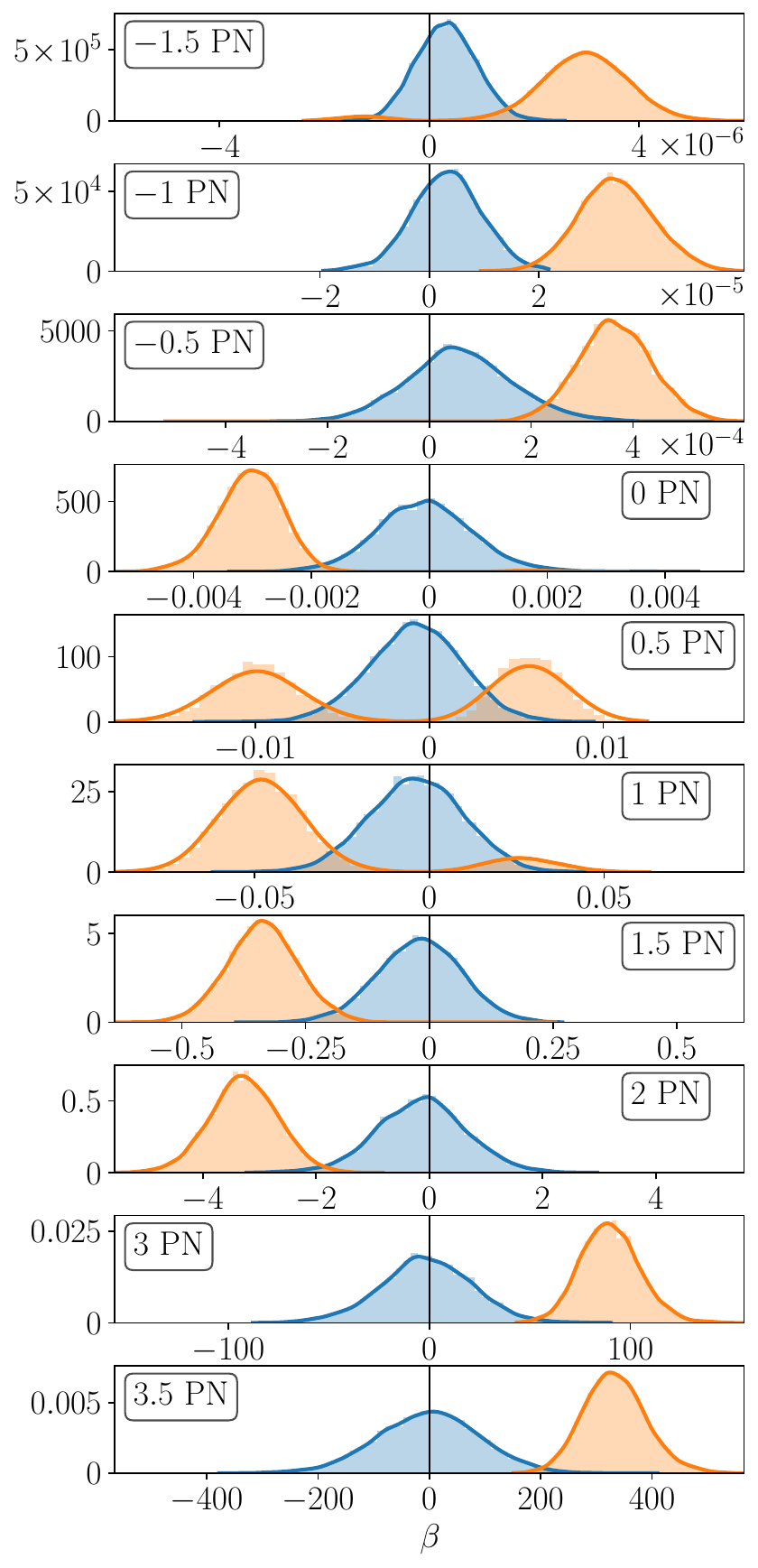}
    \caption{Marginal posterior distributions $p(\beta)$ for the ppE parameter $\beta$ at fixed network SNR $\rho=330$ for the lighter ($20\,M_\odot$, left panel) and the heavier ($60\,M_\odot$, right panel) BBH system, shown as histograms with kernel-density-estimate (KDE) overlays for ppE indices spanning $-1.5$ to $3.5$ PN order. Orange corresponds to recoveries with Model~I, while blue corresponds to recoveries with Model~II. Note that each PN order is displayed with its own $\beta$ range (and scale factor where indicated) to resolve the posterior, though all panels are aligned with $\beta = 0$ (black line). Despite $\rho$ exceeding the thresholds in Fig.~\ref{fig:threshold_snr}, Model~II remains consistent with $\beta=0$ across all PN orders. On the other hand, Model~I exhibits systematic bias, in some cases exceeding $3\sigma$. Observe also that in the heavier system (right panel) Model~I exhibits a bimodality at $0.5$ and $1$ PN order. Model~II instead remains unimodal and consistent with $\beta=0$ across all PN orders, though its median is visibly shifted toward the median of Model~I in most cases.
    }    \label{fig:ridgeline_plots}
\end{figure*}

For the heavier system, the $\log_{10}\mathrm{BF}_{\mathrm{ppE},\mathrm{GR}}$ values are scattered and mostly negative, with the exception of the $0$ and negative PN orders. Even in this case, all values are still within $\lesssim 0.5$, indicating that at $\rho=\rho_\mathrm{th}$ the false-positive GR violation \textit{survives} model selection.
Details on how the uncertainty on the Bayes factor is obtained can be found in Appendix~\ref{sec:uncertainty-bayes-factor}.

%%%%%%%%%%%%%%%%%%%%%%%%%%%%%%
\subsection{Results for recovery at high SNR}
\label{ssec:recovery-at-high-SNR}

In this subsection, we recover the injection described in Sec.~\ref{ssec:injection} using Model II ($\mathcal{H}_\mathrm{II}$) at a fixed, high network SNR of $\rho = 330$. This SNR is well above any of the thresholds reported in Fig.~\ref{fig:threshold_snr}, and therefore serves as a stringent test of whether the ppE inference with Model II remains robust in the regime where Model I is biased across all PN orders.

\begin{figure*}[htb]
    \centering
    \includegraphics[width=\columnwidth]{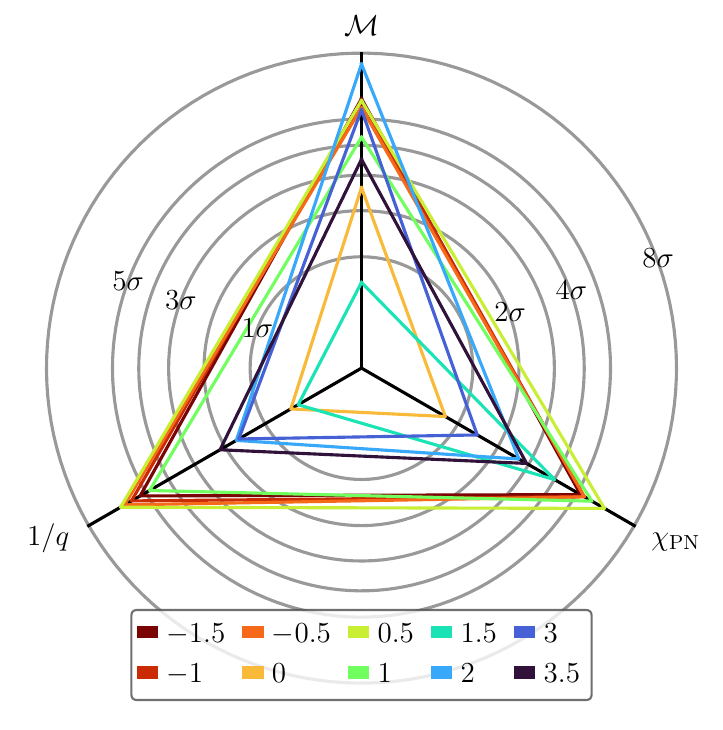}
    \includegraphics[width=\columnwidth]{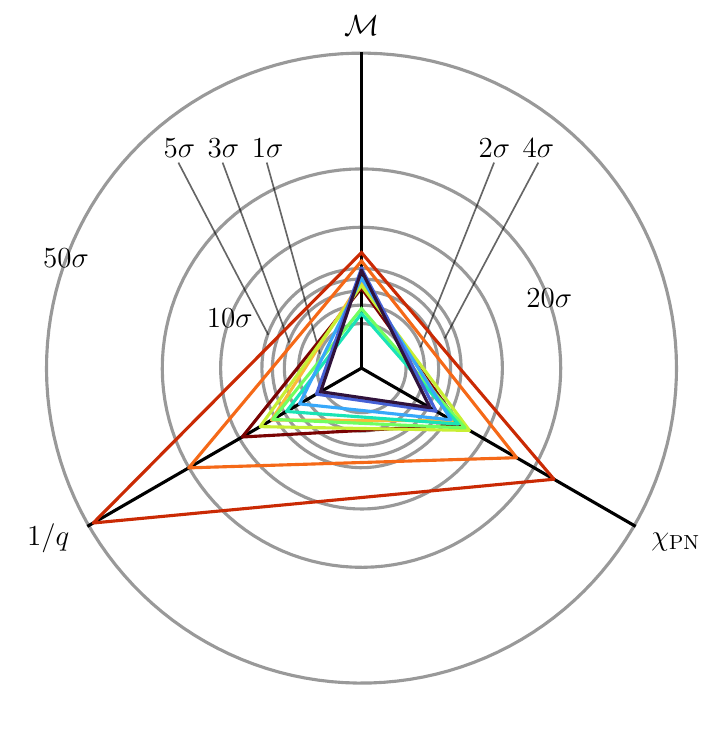}
    \caption{Radar charts summarizing the fractional systematic bias defined in Eq.~\eqref{eq:fractional-bias} found in chirp mass, inverse mass ratio, and effective spin for the $20\,M_\odot$ (left panel) and $60\,M_\odot$ (right panel) injections recovered at $\rho=330$ with Model~I. Each triangle corresponds to a different PN order at which $\beta$ enters the waveform phase (colors mapped to PN orders as indicated in the legend), and the three spokes report the bias in $\mathcal{M}$, $1/q$, and $\chi_{\mathrm{PN}}$, respectively. Concentric circles denote bias levels in units of $\sigma$. To accommodate the wide range of values, we use a nonlinear radial scaling that compresses large biases by plotting
$\sqrt{|\delta\theta|/\sigma_\theta}$ instead of $|\delta\theta|/\sigma_\theta$.}
    \label{fig:radar_charts}
\end{figure*}

The ridgeline plots in Fig.~\ref{fig:ridgeline_plots} show the marginal posteriors $p(\beta)$ recovered with Model~II against those obtained with Model~I for the $M = 20\,M_\odot$ injection (left panel) and the $M=60\,M_\odot$ injection (right panel). Each row corresponds to a fixed ppE index, spanning from $-1.5$ to $3.5$ PN order, excluding $2.5$, due to the degeneracy with $\phi_\mathrm{ref}$.

For both BBH total masses, the comparison shows that once NR-calibration uncertainty is marginalized over through a set of nuisance parameters, the inference with a ppE template no longer attributes the injection-recovery mismatch to a beyond-GR effect. Instead, the residual is accommodated within the NR-calibration degrees of freedom already allowed by the uncertainty-aware baseline.
Overall, these results indicate that incorporating NR-calibration uncertainty into the baseline waveform model is sufficient to prevent false-positive ppE tests in the high-SNR regime explored here.

Recovering at the same high SNR of $\rho = 330$ with Model~I also helps understand which recovered astrophysical parameters are more likely to be biased in response to a bias in $\beta$. To do this, we define the systematic error of a parameter $\theta$ as
\begin{equation}
\delta\theta = \theta_\mathrm{rec}-\theta_\mathrm{inj},
\end{equation}
where $\theta_\mathrm{rec}$ is the best estimate of the recovered parameter (i.e. the mean of the marginal posterior) and $\theta_\mathrm{inj}$ is its true (injected) value. It is customary to measure the magnitude of this systematic error in units of the standard deviation, $\sigma_\theta$, of its marginal posterior and define the bias as \cite{Cutler:2007mi,Littenberg:2012uj,Vallisneri:2013rc}
\begin{equation}
\text{parameter bias} := \frac{\left|\delta\theta\right|}{\sigma_\theta}.
\label{eq:fractional-bias}
\end{equation}
A value larger than unity indicates that the systematic error exceeds the statistical uncertainty, so the inference is biased at more than the $1\sigma$ level.

The radar charts in Fig.~\ref{fig:radar_charts} summarize these biases for the chirp mass $\mathcal{M}$, inverse mass ratio $1/q$, and effective spin parameter $\chi_\mathrm{PN}$, as a function of the PN order at which the ppE parameter enters the waveform phase. The left panel corresponds to the lighter ($20\,M_\odot$) BBH system, while the right panel corresponds to the heavier ($60\, M_\odot$) BBH system.
For the lighter system, most PN orders yield biases in all three parameters that exceed the $1\sigma$ level. However, when $\beta$ enters at $0$ PN order, the bias is predominantly on the $\mathcal{M}$ parameter, whereas $1/q$ and $\chi_\mathrm{PN}$ remain consistent with the injection. Similarly, at $1.5$ PN order the bias is largely carried by $\chi_\mathrm{PN}$, while $\mathcal{M}$ and $1/q$ are not significantly biased. This pattern is consistent with the interpretation that the ppE term tries to fit the injected higher-PN-order phase terms and ``drags'' the most degenerate astrophysical parameter along with it.

\begin{figure}
    \includegraphics[width=\columnwidth]{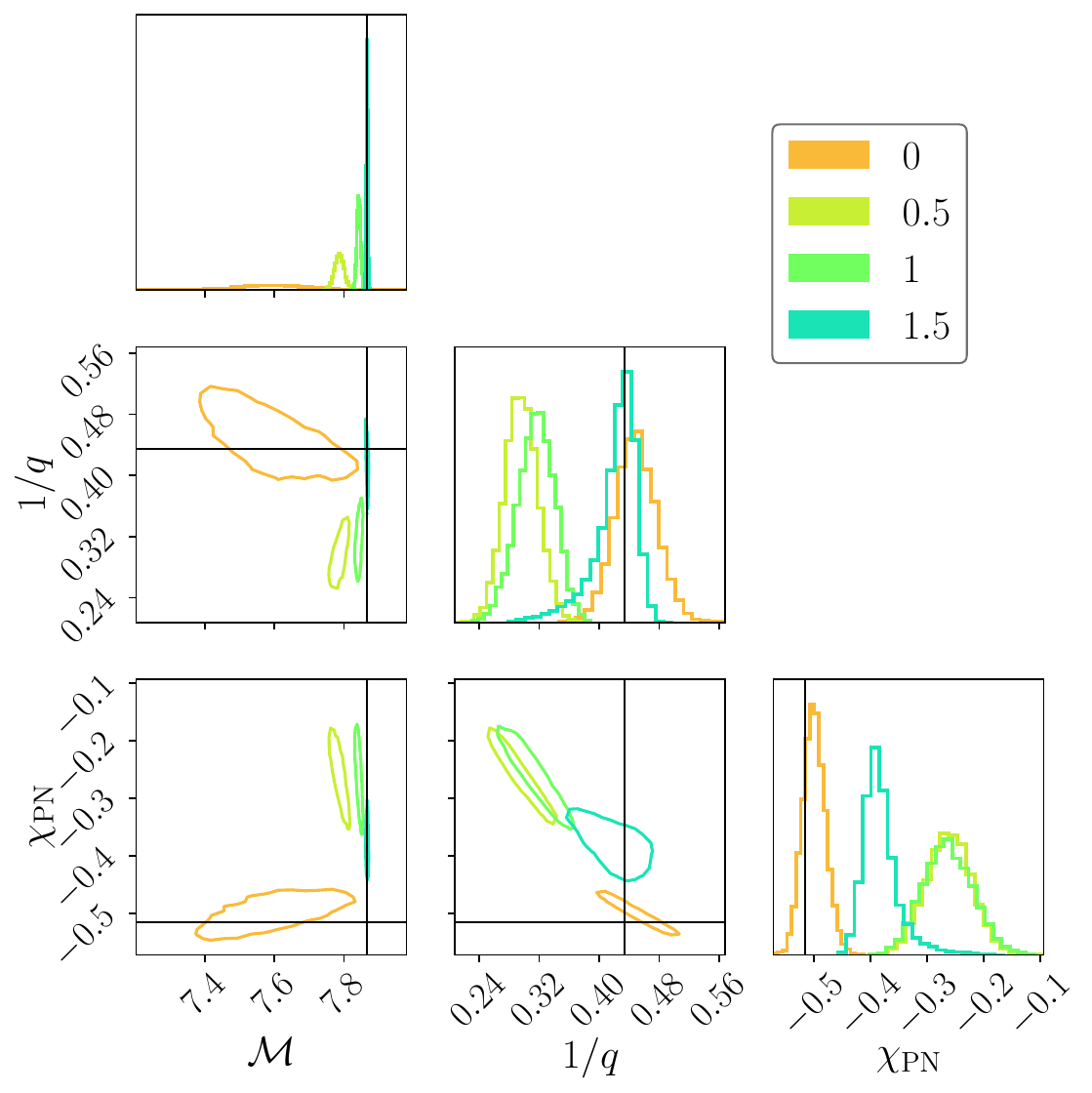}
    \caption{Corner plot showing the 1D marginals and 2D joint posteriors for $\mathcal{M}$, $1/q$, and $\chi_{\mathrm{PN}}$ for the $20\,M_\odot$ injection recovered at $\rho=330$ with Model~I, comparing runs where the ppE phase term enters at $0$, $0.5$, $1$, and $1.5$ PN order (colors as labeled). Each contour encloses the 90\% HPD credible region. Black lines mark the injected parameter values. Observe that at $0$ PN order the recovery is biased primarily in $\mathcal{M}$ but not in $1/q$ and $\chi_{\mathrm{PN}}$. Conversely, at $1.5$ PN order $\mathcal{M}$ and $1/q$ remain consistent with the injection while the dominant bias appears in $\chi_{\mathrm{PN}}$.}    \label{fig:corner_compare_PN_orders_20_MSUN}
\end{figure}

The corner plot in Fig.~\ref{fig:corner_compare_PN_orders_20_MSUN} makes this more explicit by comparing the joint posteriors of $\mathcal{M}$, $1/q$, and $\chi_\mathrm{PN}$ for $\beta$ entering at $0$, $0.5$, $1$, and $1.5$ PN order. Observe that at $0.5$ PN order the bias propagates to all astrophysical parameters as a ppE term at this order (for which the GR term is absent) can overlap with multiple GR phase contributions. In the $1$-PN-order case too, the bias affects all astrophysical parameters, and this can be interpreted as a bias first initiated in the mass ratio, which then causes a bias in the remaining parameters through their correlations.

A bias in the ``dragged'' astrophysical parameter may (as in the $1$ PN order case) or may not (as in the $0$ PN and $1.5$ PN order cases) trigger a bias in the remaining astrophysical parameters.
Similar behavior is commonly seen in GW parameter estimation. For example, in spin-aligned, highly asymmetric binaries, the phase contributions associated with the spin of the lighter object, $\chi_2$, are suppressed by the mass ratio, and therefore, even a large bias in  $\chi_2$ can have negligible impact on the other parameters.

For the heavier system, the biases are substantially larger overall and, at this SNR, all intrinsic astrophysical parameters show non-negligible biases across the full range of PN orders explored. These biases exhibit a hierarchy across PN orders: when $\beta$ enters at negative PN orders, the biases are more severe, reaching the $\sim 48\sigma$ level in $1/q$. Meanwhile, when $\beta$ enters at positive PN orders, the biases are still large but remain within $\sim 10\sigma$. Note that, in the $0.5$ and $1$ PN order cases, the marginal posteriors of the intrinsic parameters are bimodal due to a bimodality in $\beta$ (see Fig.~\ref{fig:ridgeline_plots}). In evaluating the biases, we therefore restrict to the subset of samples corresponding to $\beta\le 0$. At $\rho=330$, the two modes are clearly disjoint and each is well approximated by a Gaussian, so either branch provides a meaningful point estimate and uncertainty for the astrophysical parameters. We adopt the $\beta\le 0$ branch because it yields the smaller inferred bias.

%%%%%%%%%%%%%%%%%%%%%%%%%%%%%%%%%%%%%%%%%%%%%%%%%%%%%%%%%%%%%%%%%%%%%%%%%%%
\section{Conclusions}
\label{sec:conclusions}

In this work we examined how uncertainty in the NR-calibrated late-inspiral fitting coefficients of the phenomenological waveform model \texttt{IMRPhenomD} impacts inspiral, single-ppE parameter tests.
We illustrate, by example, how these uncertainties can bias ppE tests to yield false-positive GR deviations and investigate whether extending the GR hypothesis to include the NR-calibration uncertainty prevents these from appearing. 
We find that, when the recovery ppE template fixes the NR-calibrated fitting coefficients to their nominal \texttt{IMRPhenomD} values, a GR-consistent signal generated with an alternative but plausible set of calibration coefficients induces a nonzero ppE phase deformation at SNRs relevant for the upcoming O5 detector network.
By contrast, when using the uncertainty-aware \texttt{IMRPhenomD} baseline for the ppE test, which incorporates the NR-calibration uncertainty directly into the inference, we consistently recover a zero ppE parameter across all PN orders explored.

These results (i) motivate extra caution when interpreting positive ppE tests during future observation runs, particularly for lighter BBHs when the inferred dephasing enters at positive PN order, and for heavier BBHs when it enters at negative PN order. They also (ii) strongly support the use of uncertainty-aware waveform models in the high-SNR regime anticipated for next-generation detectors. Such models should explicitly account for NR-calibration degrees of freedom, ideally capturing more than the intrinsic NR calibration uncertainty considered here, when performing parametrized tests of GR.

Our study quantifies a clear vulnerability of gravitational-wave tests of GR in the presence of waveform-calibration systematics for two specific BBH configurations, and thus points to several natural extensions. Because we considered only two specific BBH configurations, it remains to be determined what fraction of the O5 BBH population will be susceptible to false-positive ppE inferences driven by NR-calibration systematics, and how such biases would propagate into GR tests that combine entire event catalogs. It would also be valuable to quantity the degree to which these biases persist in more recent waveform models (e.g., \texttt{IMRPhenomX}, as well as those not encompassed by the \texttt{IMRPhenom} waveform family) and, particularly, to quantify their sensitivity to alternative calibration parametrizations and NR training sets.

\begin{figure*}[htb]
    \centering
    \includegraphics[width=\columnwidth]{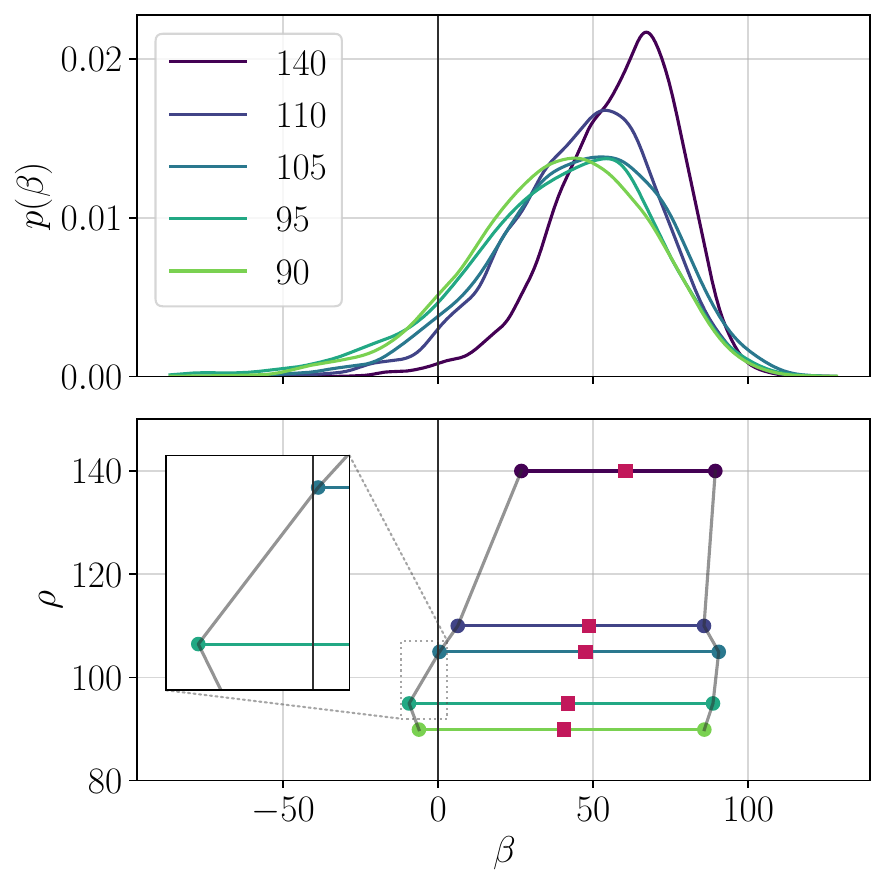}
    \includegraphics[width=\columnwidth]{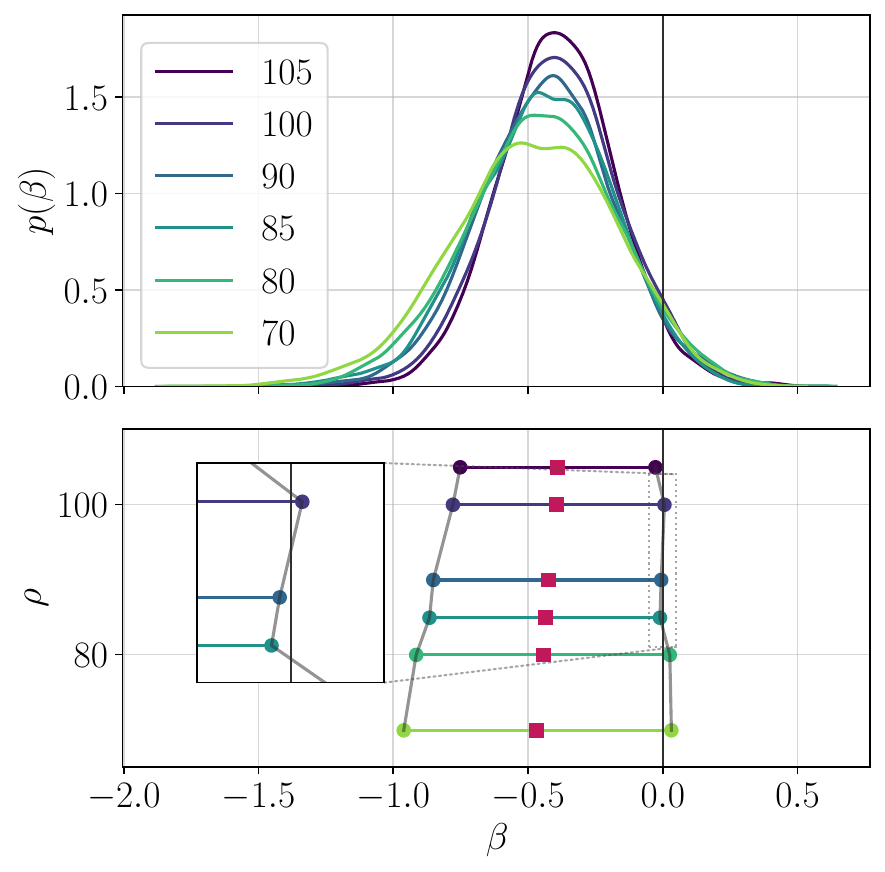}
    \caption{Bracketing of the threshold SNR for the $3$-PN-order (left panel) and $1.5$-PN-order (right panel) ppE phase term for the lighter ($20\,M_\odot$, left panel) and the heavier  ($60\,M_\odot$, right panel) binary. Top: marginal posteriors $p(\beta)$ for a sequence of network SNRs $\rho \in [90,140]$ (left panel) and $\rho=70,80,85,90,100,105$ (right panel). Bottom: corresponding 90\% HPD intervals in $\beta$ (horizontal bars) with medians shown as magenta squares. For the left panel, observe that the median drifts away from $\beta = 0$ and the threshold $\rho_{\mathrm{th}}$ is bracketed within $(95,105)$. Linear interpolation of the lower HPD bounds intersects $\beta=0$ (black vertical line) at  $\rho_{\mathrm{th}}=104.5$. For the right panel, observe that the drift of the median towards $\beta = 0$ leads to a non-monotonic crossing of $\beta = 0$, which is excluded at $\rho=85$, but is compatible again at $\rho=100$. In this case, we conservatively bracket $\rho_{\mathrm{th}}$ within $(80,105)$ and quote $\rho_{\mathrm{th}}=83.5$.
    }
    \label{fig:posterior_and_HPD_3.0PN_20_MSUN}
\end{figure*}

%%%%%%%%%%%%%%%%%%%%%%%%%%%%%%%%%%%%%%%%%%%%%%%%%%%%%%%%%%%%%%%%%%%%%%%%%%%
\section*{Acknowledgments}
We thank Rohit S. Chandramouli, Yiqi Xie, Isaac Legred, and Ania Liu for useful discussions. We thank Jocelyn~S.~Read for helpful comments.

N.Y. and S.M. acknowledge support from the Simons Foundation through Award No. 896696, the Simons Foundation International through Award No. SFI-MPS-BH-00012593-01, the NSF through Grants No. PHY-2207650 and No. PHY-25-12423, and NASA through Grant No. 80NSSC22K0806. 
C.-J.~H. acknowledges the support from the Nevada Center for Astrophysics, from NASA Grant No. 80NSSC23M0104, and the NSF through the Award No.~PHY-2409727.
We relied on a number of open-source software packages for this work, including \texttt{numpy} \cite{harris2020array}, \texttt{matplotlib} \cite{Hunter:2007}, \texttt{scipy} \cite{2020SciPy-NMeth}, \texttt{jax} \cite{jax2018github}, \texttt{flowMC} \cite{flowmc_paper}, \texttt{ripplegw} \cite{ripple_paper}, \texttt{arviz} \cite{arviz_2019}, \texttt{corner} \cite{corner}, \texttt{bilby} \cite{bilby_paper}, \texttt{parallel-bilby} \cite{pbilby_paper}, \texttt{dynesty} \cite{2020MNRAS.493.3132S,sergey_koposov_2024_12537467}, and \texttt{singularity} \cite{singularity_paper,singularity_zenodo}.

This research was supported in part by the Illinois Computes project which is supported by the University of Illinois Urbana-Champaign and the University of Illinois System.

\appendix

\section{Bracketing of the threshold SNR}
\label{sec:appendix-bracketing}
This section contains supplementary plots to detail the process of finding the threshold SNR $\rho_\mathrm{th}$.
The left panel of Fig.~\ref{fig:posterior_and_HPD_3.0PN_20_MSUN} shows the $3$ PN order case for the lighter ($20\,M_\odot$) system and provides a representative example of the SNR-increase bracketing procedure.
The upper panel shows the marginal posterior $p(\beta)$ for a sequence of network SNRs probed from $\rho = 90$ up to $\rho = 140$, using step sizes of 5 and 10 as we approach and cross the threshold. The lower panel summarizes the corresponding 90\% HPD intervals as horizontal segments (with the medians marked by magenta squares). We observe that the HPD interval shrinks with increasing $\rho$ as expected, while the median drifts approximately linearly. In this case, we bracket the threshold within the interval $(95,105)$ and obtain a best estimate of $\rho_{\mathrm{th}}$ by interpolating the lower ends of the 90\% HPD intervals and finding their intersection with $\beta=0$, which yields $\rho_{\mathrm{th}}=104.5$.

The right panel of Fig.~\ref{fig:posterior_and_HPD_3.0PN_20_MSUN} shows the $1.5$ PN order case for the heavier ($60\,M_\odot$) system and illustrates how the drift of the median towards $\beta = 0$ can complicate pinpointing $\rho_{\mathrm{th}}$.
Here we probe $\rho\in[70,105]$ in steps of 5 and 10. The asymmetric contraction of the HPD interval as $\rho$ increases causes the median to drift and, consequently, the upper HPD boundary to cross $\beta=0$ twice: $\beta=0$ is first excluded at $\rho=85$ but the HPD region becomes compatible with $\beta = 0$ again at $\rho=100$. To remain conservative in our estimate, we adopt the bracketing interval $(80,105)$ and quote $\rho_{\mathrm{th}}=83.5$ as the threshold, consistent with our definition in Sec.~\ref{ssec:threshold-SNR-and-Bayes-factor}.

\begin{figure*}[htb]
    \centering
    \includegraphics[width=\columnwidth]{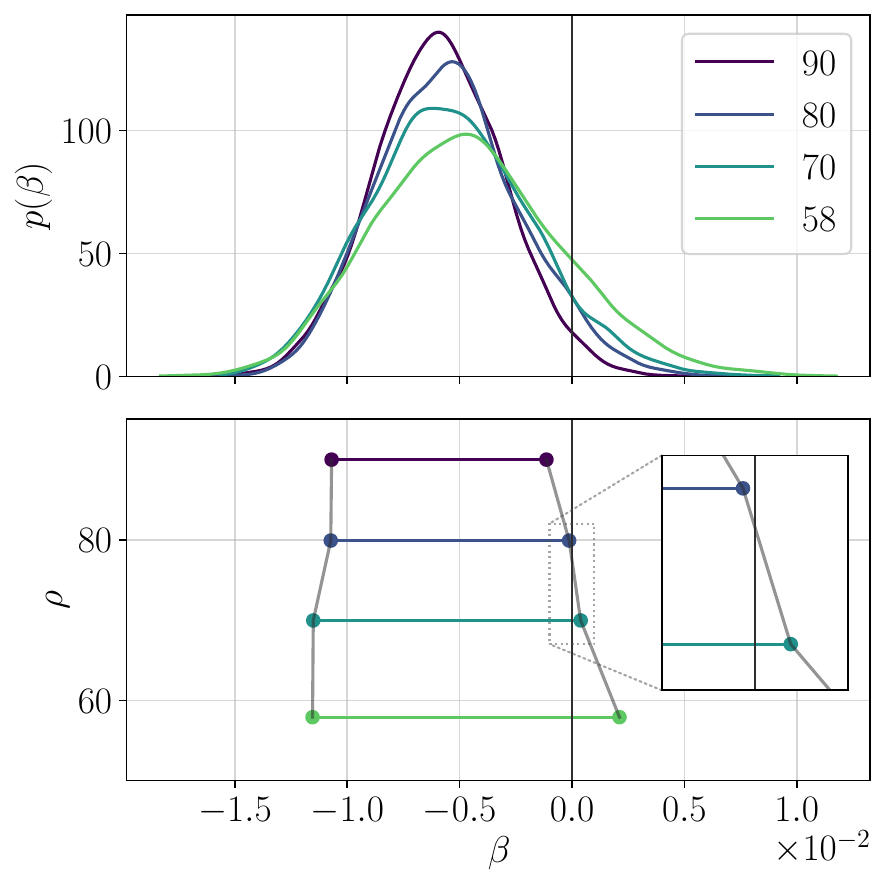}
    \includegraphics[width=\columnwidth]{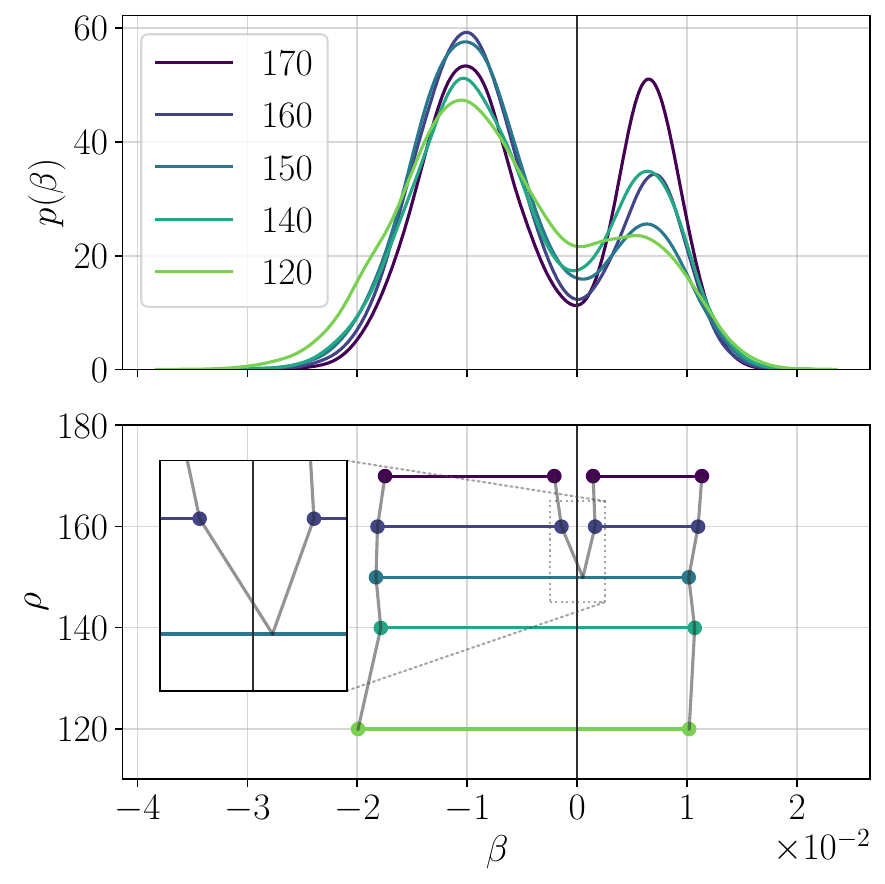}
    \caption{Bracketing of the threshold SNR for the $0.5$-PN-order ppE phase term, for the lighter ($20\,M_\odot$, left panel) and the heavier ($60\,M_\odot$, right panel) binary. Top: marginal posteriors $p(\beta)$ for the network SNRs $\rho$ probed. Bottom: corresponding 90\% HPD intervals in $\beta$ (horizontal bars). Observe that the posterior for the heavier binary develops two modes. The HPD region crosses $\beta=0$ monotonically in both cases.
    }
\label{fig:posterior_and_HPD_0.5PN_20_MSUN}
\end{figure*}
Figure~\ref{fig:posterior_and_HPD_0.5PN_20_MSUN} shows the bracketing procedure for a ppE term entering at $0.5$ PN order, for the lighter ($20 \,M_\odot$, left panel) and the heavier ($60 \,M_\odot$, right panel) system. In the lighter binary case (left panel), the marginal posterior $p(\beta)$ (top panel) remains unimodal across the SNRs probed, and the corresponding 90\% HPD intervals (bottom panel) shrink as $\rho$ increases. As in the representative $3$ PN order example (left panel of Fig.~\ref{fig:posterior_and_HPD_3.0PN_20_MSUN}), the HPD region crosses the $\beta=0$ line monotonically, so $\rho_{\mathrm{th}}$ can be bracketed cleanly between two neighboring runs. In the heavier binary case (right panel), the behavior is qualitatively different. Here, $p(\beta)$ develops two comparable modes, one at $\beta>0$ and one at $\beta<0$. As these peaks contract with increasing $\rho$, the 90\% HPD region transitions from a single connected interval to two disjoint intervals that exclude $\beta=0$. We linearly interpolate the ends of the individual HPD intervals for $\rho>150$ and we connect them to the $\rho=150$ interval at the point corresponding to the local minimum of $p(\beta)$. Like in previous cases, we determine the intersection with the line $\beta=0$, yielding $\rho_{\mathrm{th}}=152.6$, which is bracketed by the interval $(150,160)$.

%%%%%%%%%%%%%%%%%%%%%%%%%%%%%%%%%%%%%%%%
\section{Uncertainty on the Bayes factor}
\label{sec:uncertainty-bayes-factor}
In the following, we describe how we estimate the uncertainty on the Bayes factors showed in the bottom panel of Fig.~\ref{fig:threshold_snr}.
The result file of a given run provided by \texttt{bilby} contains both a value for the log-evidence $\ln Z$ and an estimate of its statistical uncertainty due to the sampling procedure via \texttt{dynesty}, $\delta \ln Z$. Taking this uncertainty at face value, the log-10 Bayes factor, $\log_{10}\mathrm{BF}_\mathrm{A,B} \pm \delta\log_{10}\mathrm{BF}_\mathrm{A,B}$, between two models, A and B, can be estimated as
\begin{align}
\log_{10}\mathrm{BF}_{\mathrm{A,B}} &= \frac{\ln Z_A - \ln Z_B}{\ln 10},\\
\delta\log_{10}\mathrm{BF}_{\mathrm{A,B}} &= \frac{\sqrt{(\delta\ln Z_A)^2 + (\delta\ln Z_B)^2}}{\ln 10}.
\end{align}
For the sampler settings we use, the typical uncertainty for the log-10 Bayes factor between $\mathcal{H}_\mathrm{I}$ and $\mathcal{H}_0$ is $\lesssim 0.1$. This is below the $\sim 0.5$ threshold that separates ``bare mention'' from stronger evidence levels on the Jeffreys' scale (Table~\ref{tab:jeffrey-scale}), so it does not obscure cases where one model is genuinely preferred over the other. Likewise, if the Bayes factor genuinely produces an inconclusive result ($|\log_{10}\mathrm{BF}_\mathrm{A,B}| <0.5$), this uncertainty is small enough to not change the outcome of the model selection.

\begin{figure}
    \centering
    \includegraphics[width=\columnwidth]{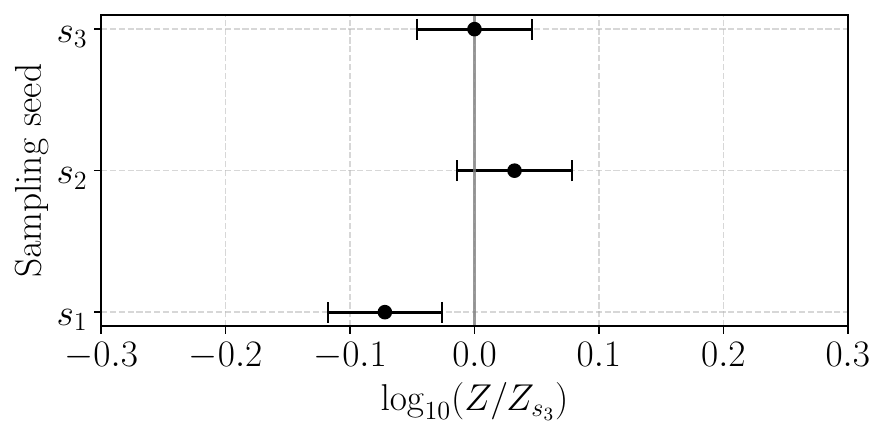}
    \caption{Consistency check of the uncertainty on the evidence reported by \texttt{bilby}. Shown are $\log_{10}(Z_{s_i}/Z_{s_3})$ for three recoveries of the $60\,M_\odot$ injection at $\rho=180$ with Model~I and a $-1.5$-PN-order ppE term, using different sampling seeds ($s_1, s_2, s_3$). The spread is consistent with the sampling uncertainty of $\lesssim 0.1$.}

\label{fig:bayes_factor_spread_-1.5PN_60_MSUN}
\end{figure}

\begin{figure}
    \centering
    \includegraphics[width=\columnwidth]{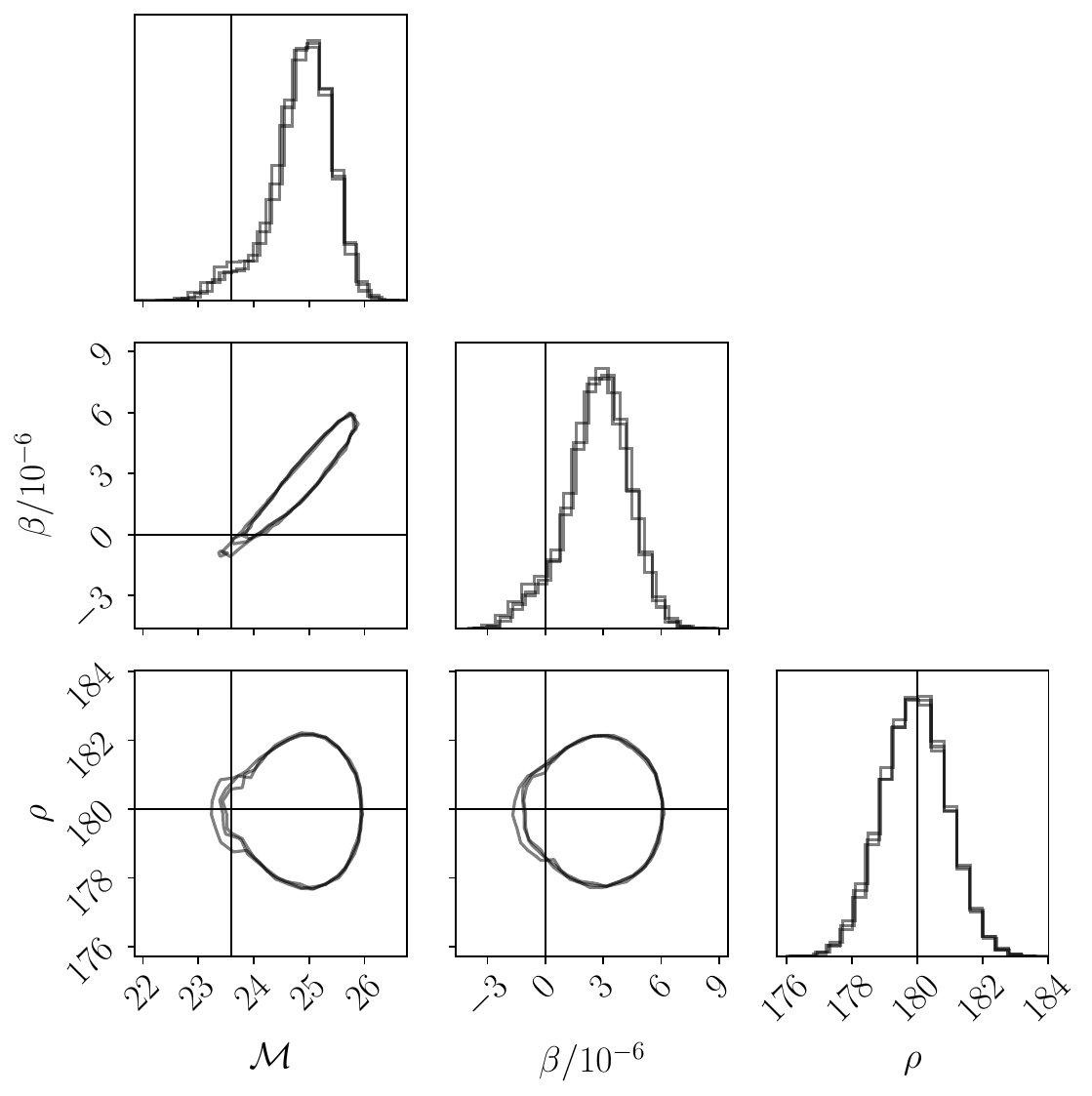}
    \caption{Corner plot comparing the posterior samples for the same $60\,M_\odot$ injection at $\rho=180$ recovered with Model~I and a $-1.5$-PN-order ppE term, using three different sampling seeds ($s_1, s_2, s_3$). Observe that the posteriors are stable with respect to a change in sampling seed.}
\label{fig:corner_different_sampling_seed_-1.5PN_60_MSUN}
\end{figure}

We validate the uncertainty on the evidence reported by \texttt{bilby} by repeating the same recovery with identical priors and high sampler settings, while varying only the random sampling seed. As an example, we inject the $60\,M_\odot$ signal at $\rho = 180$, and recover it with Model~I equipped with a ppE term entering at $-1.5$ PN order, using three different seeds ($s_1$, $s_2$, $s_3$) and \texttt{nlive} $=8000$. Figure~\ref{fig:bayes_factor_spread_-1.5PN_60_MSUN} compares the resulting evidences as Bayes factors with respect to the run with seed $s_3$, and it shows that they scatter within $\sim 0.1$. We also confirm that the posterior distributions converge to the same distribution across different sampling seeds, as shown in Fig.~\ref{fig:corner_different_sampling_seed_-1.5PN_60_MSUN}. We, therefore, take the uncertainty on the evidence provided by \texttt{bilby} as reliable and propagate it to the Bayes factors, shown in Fig.~\ref{fig:threshold_snr} with error bars.

%%%%%%%%%%%%%%%%%%%%%%%%%%%%%%%%%%%%%%%%%%%%%%%
\bibliography{paper}

\end{document}